\documentclass[
prc,
11pt,
onecolumn,
showpacs,
amsmath,
floatfix,
amssymb
]{revtex4}
\usepackage{epsfig}
\usepackage{graphicx}
\usepackage{dcolumn}
\usepackage{bm}
\def    \nn             {\nonumber}
\def\r{\mbox{{\bf  r}}}
\def\p{\mbox{\boldmath $p$}}
\def\q{\mbox{\boldmath $q$}}
\def\k{\mbox{\boldmath $k$}}
\def\t{\mbox{\boldmath $t$}}
\allowdisplaybreaks[4]

\begin{document}
\title{Quasi-elastic neutrino charged-current scattering cross sections on oxygen}
\author{A.~V.~Butkevich}
\author{S.~A.~Kulagin}
\affiliation{ Institute for Nuclear Research,
Russian Academy of Sciences,
60th October Anniversary Prosp. 7A,
Moscow 117312, Russia}
\begin{abstract}

The charged-current quasi-elastic scattering of muon neutrinos on oxygen
target is computed for neutrino energies between 200 MeV and 2.5 GeV using
the relativistic distorted-wave impulse approximation with relativistic
optical potential, which was earlier successfully applied to describe 
electron-nucleus data. 
We study both neutrino and electron processes and show that the reduced 
exclusive cross sections for neutrino and electron scattering are similar. 
The comparison with the relativistic Fermi gas model (RFGM),
which is widely used in data analyses of neutrino experiments, shows that
the RFGM fails completely when applied to exclusive cross section data and
leads to overestimated values of inclusive and total cross sections. 
We also found significant nuclear-model dependence of exclusive, inclusive 
and total cross sections for about 1~GeV energy.

\end{abstract}
 \pacs{25.30.-c 25.30.Bf, 25.30.Pt, 13.15.+g}

\maketitle

\section{Introduction}

The field of neutrino oscillations has been rapidly developing
from the observation of anomalies in cosmic rays~\cite{REV1} and solar
~\cite{REV2} neutrino data to the cross checks of these anomalies
~\cite{REV3,REV4} and most recently to terrestrial confirmations of neutrino
oscillation hypothesis (Kamland, K2K~\cite{REV5} and MINOS~\cite{REV6}). The
next steps in this field would be the precision measurements of observed mass
splitting and mixing angles
and detailed experimental study of the neutrino mixing matrix.

New, extremely intense neutrino beamlines are in operation or being
planed. The data from these experiments will greatly
increase statistics. In this situation, statistical uncertainties should be
negligible compared to systematic uncertainties (ultimate precisions).
An important source of systematic uncertainties is related to
nuclear effects in neutrino interactions. Since nuclear targets are used as
neutrino detectors, a reliable interpretation of neutrino data
requires a detailed knowledge of energy and nuclear dependence of
neutrino-nucleus ($\nu A$) cross sections.
Apparently the uncertainties in neutrino cross
sections and nuclear effects produce systematic uncertainties in the
extraction of mixing parameters.

Neutrino beams of high intensity cover the energy range from a few hundred
MeV to several GeV. In this energy regime, the dominant contribution to
neutrino-nucleus cross section comes from quasi-elastic (QE) reactions and
resonance production processes. Unfortunately, the cross section data in the
relevant energy range are rather scarce and were taken on targets that
are not used in neutrino oscillation experiments (i.e., water, iron, lead
or plastic).

A variety of  Monte Carlo codes~\cite{REV7} developed to simulate neutrino
detector response are based on a simple picture, referred to as Relativistic
Fermi Gas Model, in which the nucleus is described as a system of
quasi-free nucleons.
Comparison with high-precision electron scattering data has shown that
the accuracy of predictions of this model (inclusive cross sections) depends
significantly on momentum transfer~\cite{REV8}. For inclusive nuclear
scattering at sufficiently high momentum transfer ($\gtrsim 500$ MeV/c)
the RFGM describes general behavior of cross sections. However, the
accuracy of a Fermi gas model becomes poor as momentum transfer decreases
(see, e.g., \cite{REV9}). Furthermore, this model does not account for the
nuclear shell structure, and for this reason it fails when applied to exclusive
cross sections. There are other important effects beyond the RFGM: the final
state interaction (FSI) between the outgoing nucleon and residual nucleus and
the presence of strong short-range nucleon-nucleon ($NN$) correlations, leading
to the appearance of high-momentum and high-energy components in the
nucleon energy-momentum distribution in the target.
In the calculation of Ref.~\cite{REV10} within a plane-wave impulse
approximation (PWIA) the nucleon-nucleon correlations were included using
description of nuclear dynamics, based on nuclear many-body theory. It was
shown that the Fermi gas model overestimates the total $\nu A$ cross
section by as much as 20\% at incoming neutrino energies of about 1~GeV.
Neutral current and/or charged current (CC) neutrino-nucleus cross
sections were studied within the relativistic distorted-wave impulse
approximation (RDWIA) in Refs.~\cite{REV11,REV12,REV13,REV14} using a
relativistic shell model approach. The implementation of the final-state
interaction of the ejected nucleon has been done differently. A
description of the FSI mechanisms through the inclusion of relativistic
optical potential is presented in Refs.~\cite{REV11,REV12,REV13}. In
Refs.~\cite{REV11,REV12} important FSI effects arise from the use of
relativistic optical potential within a relativistic Green's function
approach. In Ref.~\cite{REV13}, the final state interaction was included with
and without the imaginary part of the optical potential (for inclusive cross
section). A reduction of the total cross section of at least 14\% was
found at neutrino energies of 1~GeV. The relativistic optical potential
and relativistic multiple-scattering Glauber approximation were applied in
Ref.~\cite{REV14} for the treatment of the FSI effects. Apart from
relativistic and the FSI effects. Apart from relativistic and FSI effects,
other effects may be important in neutrino-nucleus reactions. In particular,
Ref.~\cite{REV15,Kolb,Vopl,Ryck,Singh} include long-range nuclear
correlations (random-phase approximation) and FSI and Coulomb corrections 
in the calculation of $\nu {}^{12}C$ inclusive cross sections near threshold energy.

In this paper, we compute the single-nucleon knockout contribution to the
exclusive, inclusive, and total cross sections of
the charged-current QE (anti)neutrino scattering from $^{16}$O using different
approximations (PWIA and RDWIA) and the Fermi gas model. We employ the LEA
code~\cite{REV16} developed for the calculation of contribution from 1$p$- and
1$s$-state nucleons to cross sections in RDWIA. The LEA program, initially
designed for computing of exclusive proton-nucleus and electron-nucleus
scattering, was successfully tested against $A(e,e'p)$
data~\cite{REV17,REV18,REV19,REV20}, and we adopt this code for neutrino
reactions. In the PWIA, the nuclear differential cross section are
described in terms of a nuclear spectral function~\cite {Frull}, which
includes contributions from nuclear shells as well as from
the $NN$ correlations. In our approach, the effect of the $NN$ correlations in
the oxygen ground state is evaluated in the PWIA using model nucleon
high-momentum component~\cite{REV21,REV22}. We propose a way to estimate
the FSI effect on the inclusive cross sections in the presence of
short-range $NN$ correlations in the ground state. The aim of this work is
twofold. First, we compute the RDWIA CC QE neutrino cross sections.
Second, we test the RFGM against electron scattering data.

The outline of this paper is the following. In Sec.II we present the formalism
for the description of the charged-current lepton-nucleus
scattering process. The RDWIA model is briefly introduced in Sec.III. Results
of the numerical calculations are presented in Sec.IV. Our conclusions are
summarized in Sec.V.
In the appendix, we discuss the general Lorentz structure of the hadronic tensor and
give expressions for the cross sections of neutrino exclusive
scattering used in our analysis.

\section{Formalism of quasi-elastic scattering}

We consider electron and neutrino charged-current QE exclusive,
\begin{equation}\label{qe:excl}
l(k_i) + A(p_A)  \rightarrow l^{\prime}(k_f) + N(p_x) + B(p_B),      
\end{equation}
and inclusive,
\begin{equation}\label{qe:incl}
l(k_i) + A(p_A)  \rightarrow l^{\prime}(k_f) + X ,                    
\end{equation}
scattering off nuclei in a one-photon (W-boson) exchange approximation. Here $l$
labels the incident lepton [electron or muon (anti)neutrino], and
$l^{\prime}$ represents the scattered lepton (electron or muon).
Figure~1 defines our conventions for the kinematical variables, where
$k_i=(\varepsilon_i,\k_i)$ and $k_f=(\varepsilon_f,\k_f)$ are initial and
final lepton momenta, $p_A=(\varepsilon_A,\p_A)$, and
$p_B=(\varepsilon_B,\p_B)$ are the initial and final target momenta,
$p_x=(\varepsilon_x,\p_x)$ is ejectile nucleon momentum, $q=(\omega,\q)$ is
the momentum transfer carried by the virtual photon (W-boson), and
$Q^2=-q^2=\q^2-\omega^2$ is the photon (W-boson) virtuality. Normalization of
states is given by
\begin{equation}
N_i\langle p_i\vert p^{\prime}_i\rangle = 2\pi \delta^3(\p_i-\p^{\prime}_i),\nn
\end{equation}
where $N_i=m/\varepsilon$ for massive particles, or $N_i=1/{2\varepsilon}$ for
massless leptons.
\begin{figure*}
  \begin{center}
    \includegraphics[height=8cm,width=16cm]{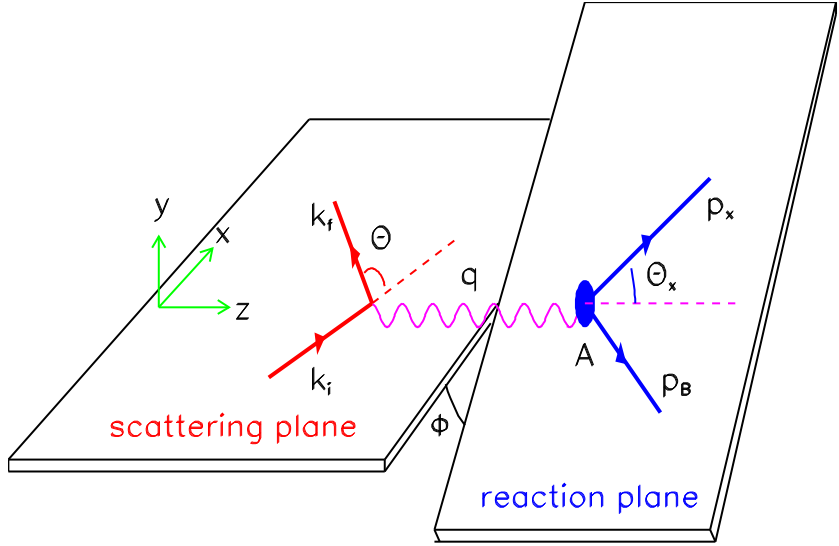}
\end{center}
\caption{(Color online) Kinematics for the quasi-elastic lepton-nucleus
  scattering process.}
\end{figure*}

\subsection{ Quasi-elastic lepton-nucleus cross sections}

In the laboratory frame, the differential cross section for exclusive
electron ($\sigma ^{el}$) and (anti)neutrino ($\sigma ^{cc}$) CC scattering can
be written as
\begin{subequations}
\begin{align}
\frac{d^6\sigma^{el}}{d\varepsilon_f d\Omega_f d\varepsilon_x d\Omega_x} &=
\frac{\vert\p_x\vert\varepsilon_x}{(2\pi)^3}\frac{\varepsilon_f}{\varepsilon_i}
 \frac{\alpha^2}{Q^4} L_{\mu \nu}^{(el)}\mathcal{W}^{\mu \nu (el)}
\\                                                                  
\frac{d^6\sigma^{cc}}{d\varepsilon_f d\Omega_f d\varepsilon_x d\Omega_x} &=
\frac{\vert\p_x\vert\varepsilon_x}{(2\pi)^5}\frac{\vert\k_f\vert}
{\varepsilon_i} \frac{G^2\cos^2\theta_c}{2} L_{\mu \nu}^{(cc)}
\mathcal{W}^{\mu \nu (cc)},
\end{align}
\end{subequations}
 where $\Omega_f$ is the solid angle for the lepton momentum, $\Omega_x$ is the
 solid angle for the ejectile nucleon momentum, $\alpha\simeq 1/137$ is the
fine-structure constant, $G \simeq$ 1.16639 $\times 10^{-11}$ MeV$^{-2}$ is
the Fermi constant, $\theta_C$ is the Cabbibo angle
($\cos \theta_C \approx$ 0.9749), $L^{\mu \nu}$ is the lepton tensor,
 $\mathcal{W}^{(el)}_{\mu \nu}$ and $\mathcal{W}^{(cc)}_{\mu \nu}$ are
correspondingly the electromagnetic and weak CC nuclear tensors which will be
 discussed below.

For exclusive reactions in which only a single discrete state or narrow
resonance of the target is excited, it is possible to integrate over the
peak in missing energy and obtain a fivefold differential cross section of
the form
\begin{subequations}
\begin{align}
\label{cs5:el}
\frac{d^5\sigma^{el}}{d\varepsilon_f d\Omega_f d\Omega_x} &= R
\frac{\vert\p_x\vert\tilde{\varepsilon}_x}{(2\pi)^3}\frac{\varepsilon_f}
{\varepsilon_i} \frac{\alpha^2}{Q^4} L_{\mu \nu}^{(el)}W^{\mu \nu (el)}
\\                                                                       
\label{cs5:cc}
\frac{d^5\sigma^{cc}}{d\varepsilon_f d\Omega_f d\Omega_x} &= R
\frac{\vert\p_x\vert\tilde{\varepsilon}_x}{(2\pi)^5}\frac{\vert\k_f\vert}
{\varepsilon_i} \frac{G^2\cos^2\theta_c}{2} L_{\mu \nu}^{(cc)}W^{\mu \nu (cc)},
\end{align}
\end{subequations}
where $R$ is a recoil factor
\begin{equation}\label{Rec}
R =\int d\varepsilon_x \delta(\varepsilon_x + \varepsilon_B - \omega -m_A)=
{\bigg\vert 1- \frac{\tilde{\varepsilon}_x}{\varepsilon_B}
\frac{\p_x\cdot \p_B}{\p_x\cdot \p_x}\bigg\vert}^{-1},                    
\end{equation}
 $\tilde{\varepsilon}_x$ is the solution to the equation
$
\varepsilon_x+\varepsilon_B-m_A-\omega=0,
$
where $\varepsilon_B=\sqrt{m^2_B+\p^2_B}$, $~\p_B=\q-\p_x$ and $m_A$ and $m_B$
are masses of the target and recoil nucleus, respectively. Note that the missing
momentum is $\p_m=\p_x-\q$.

The lepton tensor can be written as the sum of the symmetric $L^{\mu\nu}_S$ and
antisymmetric $L^{\mu\nu}_A$ tensors
\begin{subequations}
\begin{align}
\label{Lmunu}
L^{\mu\nu} &= L^{\mu\nu}_S + L^{\mu\nu}_A
\\
\label{Lmunu:S}
L^{\mu\nu}_S &= 2\left(k^{\mu}_i k^{\nu}_f + k^{\nu}_i k^{\mu}_f -
g^{\mu\nu}k_ik_f\right)\\                                                
\label{Lmunu:A}
L^{\mu\nu}_A &= h2i\epsilon^{\mu\nu\alpha\beta}(k_i)_{\alpha}(k_f)_{\beta},
\end{align}
\end{subequations}
where $h$ is $+1$ for positive lepton helicity and $-1$ for negative lepton
helicity, and $\epsilon^{\mu \nu \alpha \beta}$ is the antisymmetric tensor
with $\epsilon^{0 1 2 3}=-\epsilon_{0 1 2 3}=1$. For the scattering of
unpolarized incident electrons, $L^{\mu\nu (el)}$ only has the symmetric part
(\ref{Lmunu:S}) and the (anti)neutrino tensor $L^{\mu\nu (cc)}$ involves both
the symmetric and the antisymmetric parts.
Assuming the reference frame, in which the {\it z} axis is parallel to the
momentum transfer $\q=\k_i - \k_f$ and the {\it y} axis is
parallel to $\k_i \times \k_f$, the symmetric components
$L^{0x}_S, L^{x y}_S, L^{z y}_S$ and the antisymmetric ones $L^{0 x}_A,
L^{x z}_A, L^{0 z}_A$, as well as those obtained from them by exchanging
their indices, vanish. The electromagnetic and the weak CC hadronic tensors,
$\mathcal{W}^{(el)}_{\mu \nu}$ and $\mathcal{W}^{(cc)}_{\mu \nu}$, are given
by bilinear products of the transition matrix elements of the nuclear
electromagnetic or CC operator $J_{\mu}^{(el)(cc)}$ between the initial
nucleus state $\vert A \rangle $ and the final state $\vert B_f \rangle$ as
\begin{eqnarray}
\mathcal{W}_{\mu \nu }^{(el)(cc)} &=& \sum_f \langle B_f,p_x\vert
J^{(el)(cc)}_{\mu}\vert A\rangle \langle A\vert
J^{(el)(cc)\dagger}_{\nu}\vert B_f,p_x\rangle              
\delta (\varepsilon_A + \omega - \varepsilon_x -
\varepsilon_{B_f}),
\label{W}
\end{eqnarray}
where the sum is taken over undetected states.

In the inclusive reactions (\ref{qe:incl}) only the outgoing lepton is
detected, and the differential cross sections can be written as
\begin{subequations}
\begin{align}
\frac{d^3\sigma^{el}}{d\varepsilon_f d\Omega_f} &=
\frac{\varepsilon_f}{\varepsilon_i}
 \frac{\alpha^2}{Q^4} L_{\mu \nu}^{(el)}\overline{W}^{\mu \nu (el)},
\\                                                                       
\frac{d^3\sigma^{cc}}{d\varepsilon_f d\Omega_f } &=
\frac{1}{(2\pi)^2}\frac{\vert\k_f\vert}
{\varepsilon_i} \frac{G^2\cos^2\theta_c}{2} L_{\mu \nu}^{(cc)}
\overline{W}^{\mu \nu (cc)},
\end{align}
\end{subequations}
where $\overline{W}^{\mu \nu}$ is inclusive hadronic tensor.
A general covariant form of the hadronic tensors and the results of their
contractions with the lepton tensors are given in Appendix~\ref{A} for
exclusive lepton scattering (\ref{qe:excl}). Combining Eq.(\ref{cs5:el}) with
Eq.(A3) and Eq.(\ref{cs5:cc}) with Eq.(A7) we obtain the exclusive lepton
scattering cross sections in terms of response functions
\begin{widetext}
\begin{subequations}\label{cs5:R}
\begin{align}
\frac{d^5\sigma^{el}}{d\varepsilon_f d\Omega_f d\Omega_x} &=
\frac{\vert\p_x\vert\tilde{\varepsilon}_x}{(2\pi)^3}\sigma_M
R\big(V_LR^{(el)}_L + V_TR^{(el)}_T
 +V_{LT}R^{(el)}_{LT}\cos\phi + V_{TT}R^{(el)}_{TT}\cos 2\phi\big),
\\                                                                       
\frac{d^5\sigma^{cc}}{d\varepsilon_f d\Omega_f d\Omega_x} &=
\frac{\vert\p_x\vert\tilde{\varepsilon}_x}{(2\pi)^5}G^2\cos^2\theta_c
\varepsilon_f \vert \k_f \vert R \big \{ v_0R_0 + v_TR_T
 + v_{TT}R_{TT}\cos 2\phi + v_{zz}R_{zz}
\notag \\
& +(v_{xz}R_{xz} - v_{0x}R_{0x})\cos\phi  
-v_{0z}R_{0z} + h\big[v_{yz}(R^{\prime}_{yz}\sin\phi + R_{yz}\cos\phi)
\notag \\
& - v_{0y}(R^{\prime}_{0y}\sin\phi + R_{0y}\cos\phi) - v_{xy}R_{xy}\big]\big\},
\end{align}
\end{subequations}
\end{widetext}
where
\begin{equation}
\sigma_M = \frac{\alpha^2\cos^2 \theta/2}{4\varepsilon^2_i\sin^4 \theta/2} 
\end{equation}
is the Mott cross section. The response functions $R_i$ depend on the variables
 $Q^2,\omega,\vert\p_x \vert$, and $\theta_{x}$. Similarly, the
inclusive lepton scattering cross sections reduce to
\begin{subequations}
\begin{align}
\frac{d^3\sigma^{el}}{d\varepsilon_f d\Omega_f} &=
\sigma_M\big(V_LR^{(el)}_L + V_TR^{(el)}_T\big),
\\                                                                       
\frac{d^3\sigma^{cc}}{d\varepsilon_f d\Omega_f} &=
\frac{G^2\cos^2\theta_c}{(2\pi)^2} \varepsilon_f
\vert \k_f \vert\big ( v_0R_0 + v_TR_T
+ v_{zz}R_{zz} -v_{0z}R_{0z}- hv_{xy}R_{xy}\big),
\end{align}
\end{subequations}
where the response functions now depend only on $Q^2$ and $\omega$.

It is also useful to define a reduced cross section
\begin{equation}
\sigma_{red} = \frac{d^5\sigma}{d\varepsilon_f d\Omega_f d\Omega_x}
/K\sigma_{lN},                                                             
\end{equation}
where
$K^{el} = R {p_x\varepsilon_x}/{(2\pi)^3}$ and
$K^{cc}=R {p_x\varepsilon_x}/{(2\pi)^5}$
are phase-space factors for the electron and neutrino scattering, the recoil
factor $R$ is given by Eq.(\ref{Rec}), and $\sigma_{lN}$ is the corresponding
elementary cross section for the lepton scattering from the moving free
nucleon.

\subsection{ Nuclear current}

Obviously, the determination of the response tensor $W^{\mu\nu}$ requires the
knowledge of the nuclear current matrix elements
in Eq.(\ref{W}). We describe the lepton-nucleon scattering in the impulse
approximation (IA), assuming that the incoming lepton interacts with only one
nucleon, which is subsequently emitted. The nuclear current is written as the
sum of single-nucleon currents. Then, the nuclear matrix element
in Eq.(\ref{W}) takes the form
\begin{eqnarray}
\langle p,B\vert J^{\mu}\vert A\rangle &=& \int d^3r~ \exp(i\t\cdot\r)
\overline{\Psi}^{(-)}(\p,\r)
\Gamma^{\mu}\Phi(\r),                                                     
\end{eqnarray}
where $\Gamma^{\mu}$ is the vertex function, $\t=\varepsilon_B\q/W$ is the
recoil-corrected momentum transfer, $W=\sqrt{(m_A + \omega)^2 - \q^2}$ is the
invariant mass, $\Phi$ and $\Psi^{(-)}$ are relativistic bound-state and
outgoing wave functions.

For electron scattering, most calculations use the CC2 electromagnetic vertex
function for a free nucleon \cite{REV26}
\begin{equation}
\Gamma^{\mu} = F^{(el)}_V(Q^2)\gamma^{\mu} + {i}\sigma^{\mu \nu}\frac{q_{\nu}}
{2m}F^{(el)}_M(Q^2),                                                      
\end{equation}
where $\sigma^{\mu \nu}=i[\gamma^{\mu}, \gamma^{\nu}]/2$, $F^{(el)}_V$ and
$F^{(el)}_M$ are the Dirac and Pauli nucleon form factors. Because the bound
nucleons are off shell,
the vertex $\Gamma^{\mu}$ in Eq.(13) should be taken for the off-shell region.
We employ the de Forest prescription for off-shell vertex~\cite{REV26}
\begin{equation}
\tilde{\Gamma}^{\mu} = F^{(el)}_V(Q^2)\gamma^{\mu} + {i}\sigma^{\mu \nu}
\frac{\tilde{q}_{\nu}}{2m}F^{(el)}_M(Q^2),                                
\end{equation}
where $\tilde{q}=(\varepsilon_x - \tilde{E},\q)$ and the nucleon energy
$\tilde{E}=\sqrt{m^2+(\p_x - \q)^2}$ is placed on shell. We use the
approximation of~\cite{REV27} on the nucleon form factors. The Coulomb gauge
is assumed for the single-nucleon current.

The single-nucleon charged current  has the $V{-}A$
structure $J^{\mu (cc)} = J^{\mu}_V + J^{\mu}_A$. For a free nucleon vertex
function $\Gamma^{\mu (cc)} = \Gamma^{\mu}_V + \Gamma^{\mu}_A$, we use the CC2
vector current vertex function
\begin{equation}
\Gamma^{\mu}_V = F_V(Q^2)\gamma^{\mu} + {i}\sigma^{\mu \nu}\frac{q_{\nu}}
{2m}F_M(Q^2),                                                            
\end{equation}
and the axial current vertex function
\begin{equation}
\Gamma^{\mu}_A = F_A(Q^2)\gamma^{\mu}\gamma_5 + F_P(Q^2)q^{\mu}\gamma_5.  
\end{equation}
The weak vector form factors $F_V$ and $F_M$ are related to the corresponding
electromagnetic ones for proton $F^{(el)}_{i,p}$ and neutron $F^{(el)}_{i,n}$
by the hypothesis of the conserved vector current (CVC)
\begin{equation}
F_i = F^{(el)}_{i,p} - F^{(el)}_{i,n}.                                 
\end{equation}
The axial $F_A$ and psevdoscalar $F_P$ form factors in the dipole
approximation are parameterized as
\begin{equation}
F_A(Q^2)=\frac{F_A(0)}{(1+Q^2/M_A^2)^2},\quad                          
F_P(Q^2)=\frac{2m F_A(Q^2)}{m_{\pi}^2+Q^2},
\end{equation}
where $F_A(0)=1.267$, $m_{\pi}$ is the pion mass, and $M_A\simeq 1.032$ GeV is
the axial mass. We use the de Forest prescription for off-shell
extrapolation of $\Gamma^{\mu (cc)}$. Similar to the electromagnetic current,
the Coulomb gauge is applied for the vector current $J_V$.

\section{Model}

In Ref.~\cite{REV28}, a formalism was developed for the $A(\vec{e},e^{\prime}\vec{N})B$ reaction
that describes channel coupling in the FSI of the $N+B$ system.
According to Ref.~\cite{REV28}, a projection operator $P$ for model space was
introduced. In the independ particle shell model (IPSM), the model space for
$^{16}$O$(e,e^{\prime}N)$ consists of $1s_{1/2}$, $1p_{3/2}$, and $1p_{1/2}$
nucleon-hole states in $^{15}$N and $^{16}$O nuclei, for a total of six states.
The $1s_{1/2}$ state is regarded as a discrete state even though its spreading
width is actually appreciable. For single nucleon knockout, the parentage
expansion of the target ground-state can be written as
\begin{equation}
P\Psi_0 = \sum_{\beta \gamma}c_{\beta \gamma}\phi_{\beta \gamma}
\Phi_{\gamma},                                                           
\end{equation}
where $c_{\beta \gamma}$ is a parentage coefficient and $\phi_{\beta \gamma}$
is an overlap wave function for removal of a nucleon with single-particle
quantum number $\beta$ while leaving the residual nucleus in the state
$\Phi_{\gamma}$.
Assuming that the overlap wave functions are described by the
Dirac equation, they can be represented by a Dirac spinor of the form
\begin{equation}
\phi_{\beta \gamma}
=
\left(
\begin{array}{r}
F_{\beta \gamma} \\                                                     
i G_{\beta \gamma}
\end{array} \right)~.
\end{equation}
Similarly, for the scattering state
\begin{equation}
P\Psi^{(+)}_{\alpha} = \sum_{\beta}\psi^{(+)}_{\alpha \beta}
\Phi_{\beta}                                                           
\end{equation}
is an incoming wave function of the $N+B$ system containing an incident
plane wave in the channel $\alpha$ and outgoing spherical waves in all open
channels $\beta$ for $B(N,N^{\prime})B^{\prime}$ reaction. The Dirac
representation of distorted spinor wave functions is
\begin{equation}
\psi^{(+)}_{\alpha\beta}
=N_{\alpha}
\left(
\begin{array}{r}
\chi_{\alpha \beta } \\                                                 
i \zeta_{\alpha \beta}
\end{array} \right)~,
\end{equation}
where
\begin{equation}
N_{\alpha}=\sqrt{\frac{E_{\alpha}+m}{2E_{\alpha}}}                       
\end{equation}
is the asymptotic wave function for channel $\alpha$ normalized to unit flux, and
$E_{\alpha}=\sqrt{k^2_{\alpha}+m^2}$ is the channel energy in the
barycentric frame (the rest frame of residual nucleus $B$).

Working in coordinate space, we can write the matrix elements of the current
operator (16) for single-nucleon knockout leaving the residual nucleus in
asymptotic channel $\alpha$ as follows
\begin{eqnarray}
\langle p,B_{\alpha} \vert J^{\mu}\vert A\rangle &=& \sum_{\beta\gamma m_{b}
m^{\prime}_{b}} c_{\beta \gamma} \int d^3r \exp(i\t\cdot\r)
\langle\bar{\psi}^{(-)}_{\alpha\beta}
\vert\r m_b\rangle\nn \\
& & \times \langle\r m_b\vert\tilde{\Gamma}^{\mu}\vert\r m^{\prime}_b\rangle
\langle\r m^{\prime}_b\vert \phi_{\beta\gamma}\rangle.
\end{eqnarray}
Matrix elements of the single-nucleon current can be expressed in the
block-matrix form
\begin{equation}
\tilde{\Gamma}^{\mu}
=
\left(
\begin{array}{r}
\begin{array}{rr}
\tilde{\Gamma}^{\mu}_{++} & \tilde{\Gamma}^{\mu}_{+-} \\                  
\tilde{\Gamma}^{\mu}_{-+} & \tilde{\Gamma}^{\mu}_{--}
\end{array}
\end{array} \right)~,
\end{equation}
where each of the elements
$\langle\r m_b\vert \tilde{\Gamma}^{\mu}_{\lambda \lambda^{\prime}}
\vert \r m^{\prime}_b \rangle$ is a $2 \times 2$ spin matrix, while
$\lambda=\{+,-\}$ and $\lambda^{\prime}=\{+,-\}$ are for the upper $(+)$ and
lower $(-)$ Dirac components. Let
\begin{equation}
\langle \r m^{\prime}_b\vert\phi_{\beta \gamma}\rangle
=
\left(
\begin{array}{r}
F_{\beta \gamma m^{\prime}_b(\r)} \\                                    
i G_{\beta \gamma m^{\prime}_b(\r)}
\end{array} \right)~
\end{equation}
be the bound state overlap wave function and
\begin{equation}
\langle\bar{\psi}^{(-)}_{\alpha\beta}\vert\r m_b\rangle
=N_{\alpha}
\left(
\begin{array}{r}
\chi^{(-)\ast}_{\alpha \beta m_b(\r)} \\                                
-i \zeta^{(-)\ast}_{\alpha \beta m_b(\r)}
\end{array} \right)~
\end{equation}
be the Dirac adjoint of time-reversed distorted waves.

For the sake of application to cross section calculations, we consider the
relativistic bound-state functions within the
Hartree--Bogolioubov approximation in the $\sigma$-$\omega$ model~\cite{REV29}. 
In the mean-field approximation, the meson field
operators are replaced by their expectation values. The upper and lower radial
wave functions in the partial-wave expansion for bound-state wave functions
satisfy the usual coupled differential equations
\begin{subequations}
\begin{align}
\bigg(\frac{d}{dr} + \frac{\kappa_{\gamma}+1}{r}\bigg)F_{\beta \gamma}(r)&=
 \big[E_{\gamma} + m
+S_{\gamma}(r) - V_{\gamma}(r)\big]G_{\beta \gamma}(r),
\\                                                                       
\bigg(\frac{d}{dr} - \frac{\kappa_{\gamma}+1}{r}\bigg)G_{\beta \gamma}(r)&=
 \big[-E_{\gamma} + m
+S_{\gamma}(r) + V_{\gamma}(r)\big]F_{\beta \gamma}(r),
\end{align}
\end{subequations}
where $S_{\gamma}$ and $V_{\gamma}$ are spherical scalar and vector
potentials, and $j_{\gamma}=\vert \kappa_{\gamma}\vert-1/2$ is the total angular
momentum. Note that these potentials generally depend on the state of
the residual nucleus that is marked by subscript $\gamma$. The radial wave
functions are normalized as
\begin{equation}
\int dr~r^2 \left({\vert F_{\beta \gamma}\vert}^2 +                      
{\vert G_{\beta \gamma}\vert}^2 \right) =1.
\end{equation}
The missing momentum distribution is determined by the wave functions in
momentum space
\begin{subequations}
\begin{align} \label{FG:p}
\tilde{F}_{\beta \gamma}(p) &=
 \int dr~r^2 j_{l_{\gamma}}(p r)F_{\beta \gamma}(r),
\\                                                                       
\tilde {G}_{\beta \gamma}(p) &=
 \int dr~r^2 j_{l^{\prime}_{\gamma}}(p r) G_{\beta \gamma}(r),
\end{align}
\end{subequations}
where $j_l(x)$ is the Bessel function of order $l$ and
$l^{\prime}_{\gamma}=2j_{\gamma}-l_{\gamma}$.
If only a single state of residual nucleus is considered, or if
relativistic potentials $S$ and $V$ weakly depend on the state $\gamma$ of
residual nucleus, the relativistic momentum distribution can be written
in terms of Eq.(30) as
\begin{equation} \label{momdis:1}
P_{\beta}(p_m) = \frac{\vert c_{\beta}\vert^2}{2\pi^2}
\left(\vert \tilde{F}_{\beta}(p_m)\vert^2 +                              
\vert\tilde{G}_{\beta}(p_m)\vert^2 \right).
\end{equation}

In this work, the current operator CC2 and the bound-nucleon wave functions
~\cite{REV30} (usually referred to as NLSH) are used in the numerical analysis.
Note that the calculation of the bound-nucleon wave function for $1p_{3/2}$ 
state includes the incoherent contribution of the unresolved
$2s_{1/2}d_{5/2}$ doublet. The wave functions for these states were taken from
the parameterization of Ref.~\cite{REV31}. We use also the following values of
normalization factors $S_{\alpha}=\vert c_{\alpha}\vert^2$ relative to the full
occupancy of $^{16}$O:
$S(1p_{3/2})=0.66$,
$S(1p_{1/2})=0.7$~\cite{REV19}, and $S(1s_{1/2})=1$.

The distorted wave functions are evaluated using a relativized
Schr\"{o}dinger equation for upper components of Dirac wave functions. For
simplicity, we consider a single-channel Dirac equation
\begin{equation}
\left[{\bf \alpha}\cdot\p + \beta(m + S)\right]\psi = (E - V)\psi,       
\end{equation}
where
\begin{equation}
\psi(\r)
=
\left(
\begin{array}{r}
\psi_{+}(\r) \\                                                         
\psi_{-}(\r)
\end{array} \right)~
\end{equation}
is the four-component Dirac spinor. Using the direct Pauli reduction
method~\cite{REV32,REV33}, the system of two coupled first-order
radial Dirac equations can be reduced to a single second-order equation
\begin{equation}
\left[\nabla^2 + k^2 -2\mu\left(U^C + U^{LS}{\bf L}\cdot{\bf \sigma}     
\right)\right]\xi = 0,
\end{equation}
where $\xi$ is a two-component Pauli spinor. Here $k$ is the relativistic wave
number, $\mu$ is the reduced mass of the scattering state, and
\begin{subequations}
\begin{align}
U^C &= \frac{E}{\mu}\bigg[V + \frac{m}{E}S + \frac{S^2-V^2}{2E}\bigg]
+ U^D,
 \\                                                                      
U^D &= \frac{1}{2\mu}\bigg[-\frac{1}{2r^2D}\frac{d}{dr}
\big(r^2D^{\prime}\big)+
\frac{3}{4}\bigg(\frac{D^{\prime}}{D}\bigg)^2\bigg],
 \\
U^{LS} &= -\frac{1}{2\mu r}\frac{D^{\prime}}{D},
 \\
D &= 1 + \frac{S-V}{E + m}.
\end{align}
\end{subequations}
where $D'=dD/dr$, and $D(r)$
is known as the Darwin nonlocality factor, and $U^C$ and $U^{LS}$ are
the central and spin-orbit potentials. The upper and lower components of the
Dirac wave functions are then obtained using
\begin{subequations}
\begin{align}
\psi_{+} &= D^{1/2}\xi,
 \\                                                                      
\psi_{-} &= \frac{\bm{\sigma}\cdot \p}{E + m + S -V}~\psi_{+}.
\end{align}
\end{subequations}
Assuming a similar relationship for the coupled-channel case, i.e.,
\begin{equation}
\zeta^{(+)}_{\alpha \beta}(\r) = \frac{{\bf \sigma}\cdot \p}{E_{\beta} + m +
S_{\beta} -V_{\beta}}~\chi^{(+)}_{\alpha \beta}(\r),                     
\end{equation}
the lower components of the radial wave functions in the partial-wave expansion
for distorted waves (31) can be approximated as
\begin{equation}
\zeta^{(+)}_{\alpha \beta}(r) = \left(E_{\beta} + m +
S_{\beta} -V_{\beta}\right)^{-1}\bigg(\frac{d}{dr}+
\frac{\kappa_{\beta}}{r}\bigg)\chi^{(+)}_{\alpha \beta}(r).               
\end{equation}
We use the LEA program \cite{REV16} for the numerical calculation of the
distorted wave functions with the EDAD1 SV relativistic optical potential
\cite{REV34}. This code employs an iteration algorithm to solve the relativized
Schr\"{o}dinger equation.

A complex relativistic optical potential with a nonzero imaginary part
generally produces an absorption of flux. For the exclusive channel, this reflects
the coupling between different open reaction channels. However, for
the inclusive reaction the total flux must conserve. Currently there is no
fully consistent solution to this problem, and different approaches are
used. The Green's function approach, where the FSI effect in inclusive
reactions is treated by means of a complex optical potential and the total
flux is conserved, is presented in Refs.\cite{REV11,REV35}. To
demonstrate the effect of the optical potential on the inclusive
reactions, the results obtained in this approach were compared with those
obtained with the same potential but with the imaginary part set to 0.
It was shown that the inclusive CC neutrino cross sections calculated with
only the real part of optical potential are almost identical to those
of the Green's function approach~\cite{REV11,REV12}. A similar
approximation was used also in Ref.~\cite{REV13} to study the FSI effect
on the inclusive cross section.
In this work, in order to calculate the inclusive and total cross sections,
we use the approach in which only the real part of the optical potential
EDAD1 is included. Then the contribution of the $1p$ and $1s$ states to the
inclusive cross section can be obtained by integrating the exclusive
cross sections (11) over the azimuthal angle $\phi$ and missing momentum, that is,
$p_m$
\begin{eqnarray}
\bigg(\frac{d^3\sigma}{d\varepsilon_f d\Omega_f}\bigg)_{\rm RDWIA} &=&
\int_{0}^{2\pi}d\phi\int_{p_{min}}^{p_{max}}
dp_m\frac{p_m}{p_x \vert\q \vert}R_c\nn
\times\bigg(\frac{d^5\sigma}{d\varepsilon_f d\Omega_f d\Omega_x}\bigg)_{\rm RDWIA},
\end{eqnarray}
where $p_m=\vert\p_m\vert,~ p_x=\vert\p_x\vert,~ \p_m=\p_x-\q$, and
\begin{subequations}
\begin{align}
\cos\theta_{x} &= \frac{\p^2_x + \q^2 - \p^2_m}{2p_x\vert\q\vert},    
\\
R_c &= 1 + \frac{\varepsilon_x}{2p^2_x\varepsilon_B}
(\p^2_x +\q^2 - \p^2_m).
\end{align}
\end{subequations}
The integration limits $p_{min}$ and $p_{max}$ are given in Ref.~\cite{REV22}.
The effect of the FSI on the inclusive cross section can be evaluated using
the ratio
\begin{equation} \label{Lambda-factor}
\Lambda(\varepsilon_f,\Omega_f) =
\bigg(\frac{d^3\sigma}{d\varepsilon_f d\Omega_f}\bigg)_{\rm RDWIA}\bigg/      
\bigg(\frac{d^3\sigma}{d\varepsilon_f d\Omega_f}\bigg)_{\rm PWIA},
\end{equation}
where $\left(d^3\sigma/d\varepsilon_f d\Omega_f\right)_{PWIA}$ is the result
obtained in the PWIA.

According to data from the Thomas Jefferson National Accelerator Facility
(JLab) \cite{REV19}, the occupancy of the IPSM orbitals of
 $^{16}$O is approximately 75\% on average. In this paper, we assume that the
missing strength can be attributed to the short-range $NN$ correlations in the
ground state. To estimate this effect in the inclusive cross sections, 
we consider a phenomenological model. This model
incorporates both the single particle nature of the nucleon
spectrum at low energy and high-energy and high-momentum components due to
$NN$ correlations. The high-momentum part $P_{\rm HM}$ of the spectral function is
determined by excited states with one or more nucleons in continuum.
The detailed description of this model is given in Refs.\cite{REV21,REV22}.

In our calculations the spectral function $P_{\rm HM}$ incorporates 25\% of the
total normalization of the spectral function. The FSI effect for
the high-momentum component is estimated by scaling the PWIA result
$(d^3\sigma/d\varepsilon_f d\Omega_f)_{HM}$ with $\Lambda(\varepsilon_f,
\Omega_f)$ function (\ref{Lambda-factor}). Then the total inclusive cross
section can be written as
\begin{equation}
\frac{d^3\sigma}{d\varepsilon_f d\Omega_f} =
\bigg(\frac{d^3\sigma}{d\varepsilon_f d\Omega_f}\bigg)_{\rm RDWIA} +
\Lambda(\varepsilon_f, \Omega_f)\bigg(\frac{d^3\sigma}                 
{d\varepsilon_f d\Omega_f}\bigg)_{\rm HM}.
\end{equation}
More details about calculation of the
$(d^3\sigma/d\varepsilon_f d\Omega_f)_{\rm HM}$ can be found in Ref.~\cite{REV8}.
\begin{figure*}
  \begin{center}
    \includegraphics[height=16cm,width=16cm]{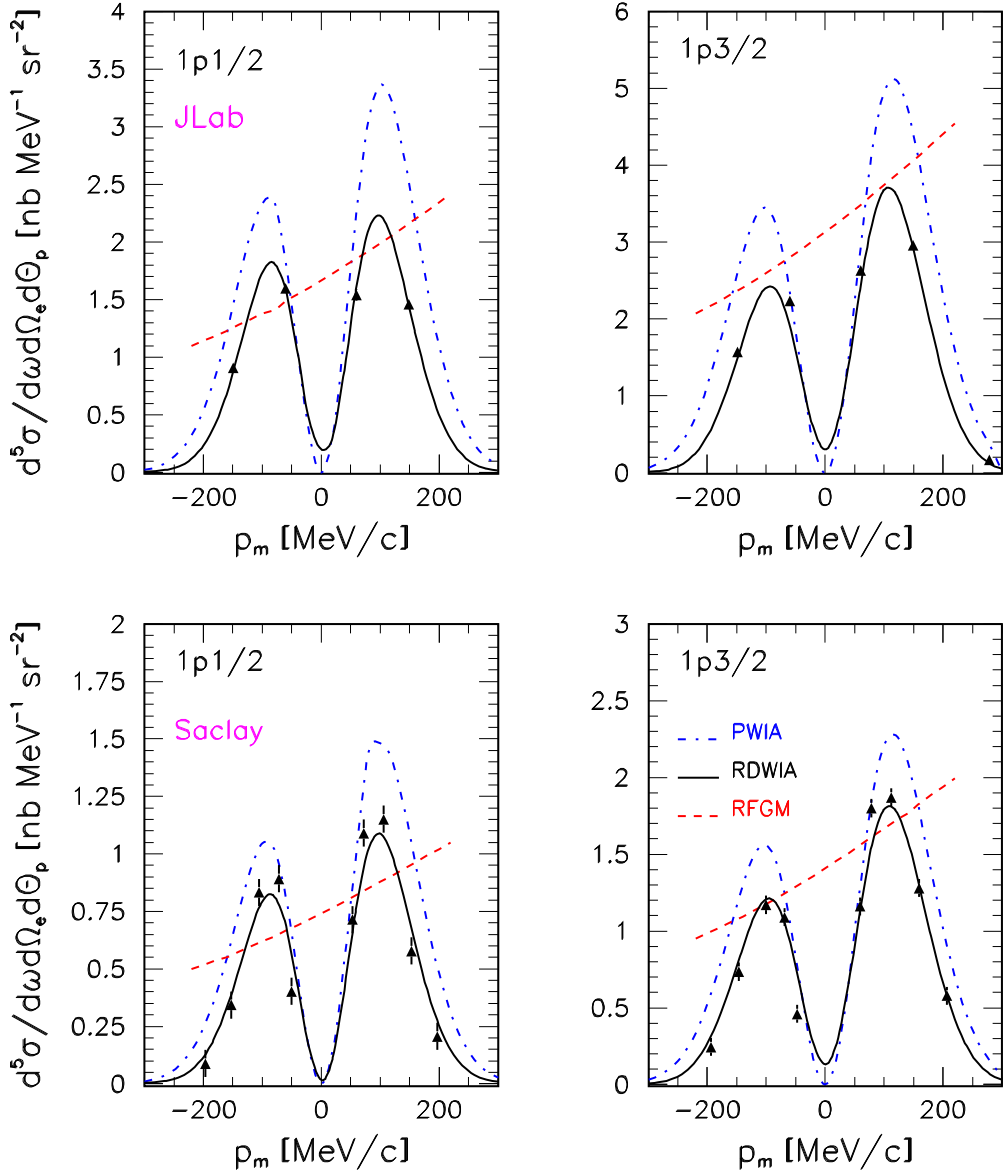}
\end{center}
\caption{(Color online) 
Calculations compared with measured differential exclusive cross section data for the
removal of protons from the 1$p$ shell of $^{16}$O as a function of missing
momentum. Upper panels: JLab data \cite{REV19} for beam energy
$E_{\rm beam}$=2.442 GeV, proton kinetic energy $T_p$=427 MeV, and $Q^2$=0.8
GeV$^2$. Lower panels: Saclay data \cite{REV36} for $E_{\rm beam}$=580 MeV,
$T_p$=160 MeV, and $Q^2$=0.3 GeV$^2$.
}
\end{figure*}
\begin{figure*}
\begin{center}
\includegraphics[height=16cm,width=16cm]{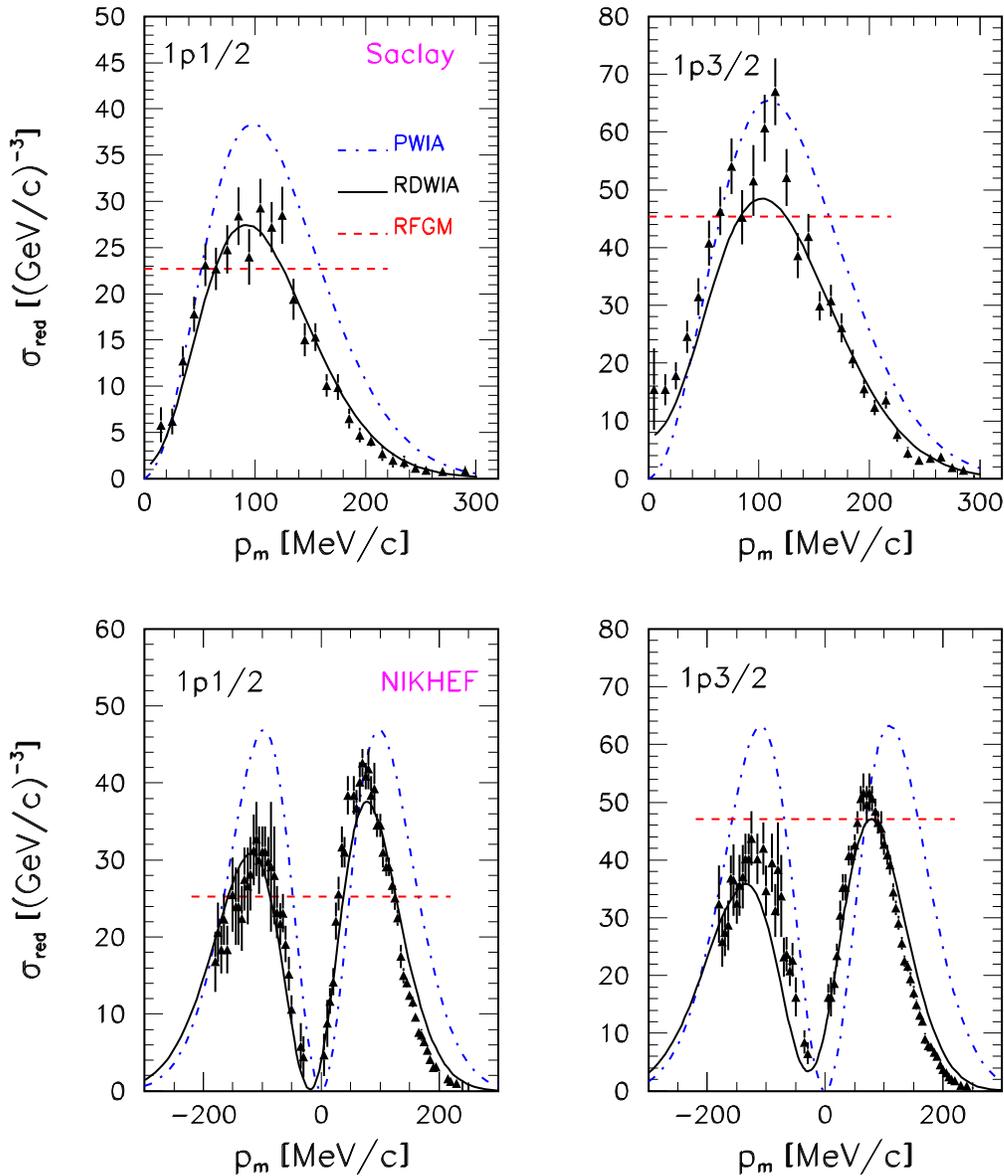}  
\end{center}
\caption{(Color online) 
Calculations compared with measured reduced exclusive cross section data for the
removal of protons from the 1$p$ shell of $^{16}$O as a function of missing
momentum.  Upper panels: Saclay data \cite{REV37} for beam energy
$E_{\rm beam}$=500 MeV, proton kinetic energy $T_p$=100 MeV, and $Q^2$=0.3
GeV$^2$. Lower panels: NIKHEF data \cite{REV38} for $E_{\rm beam}$=521
MeV, $T_p$=96 MeV, and $Q^2$ is varied. 
}
\end{figure*}
\begin{figure*}
  \begin{center}
    \includegraphics[height=16cm,width=16cm]{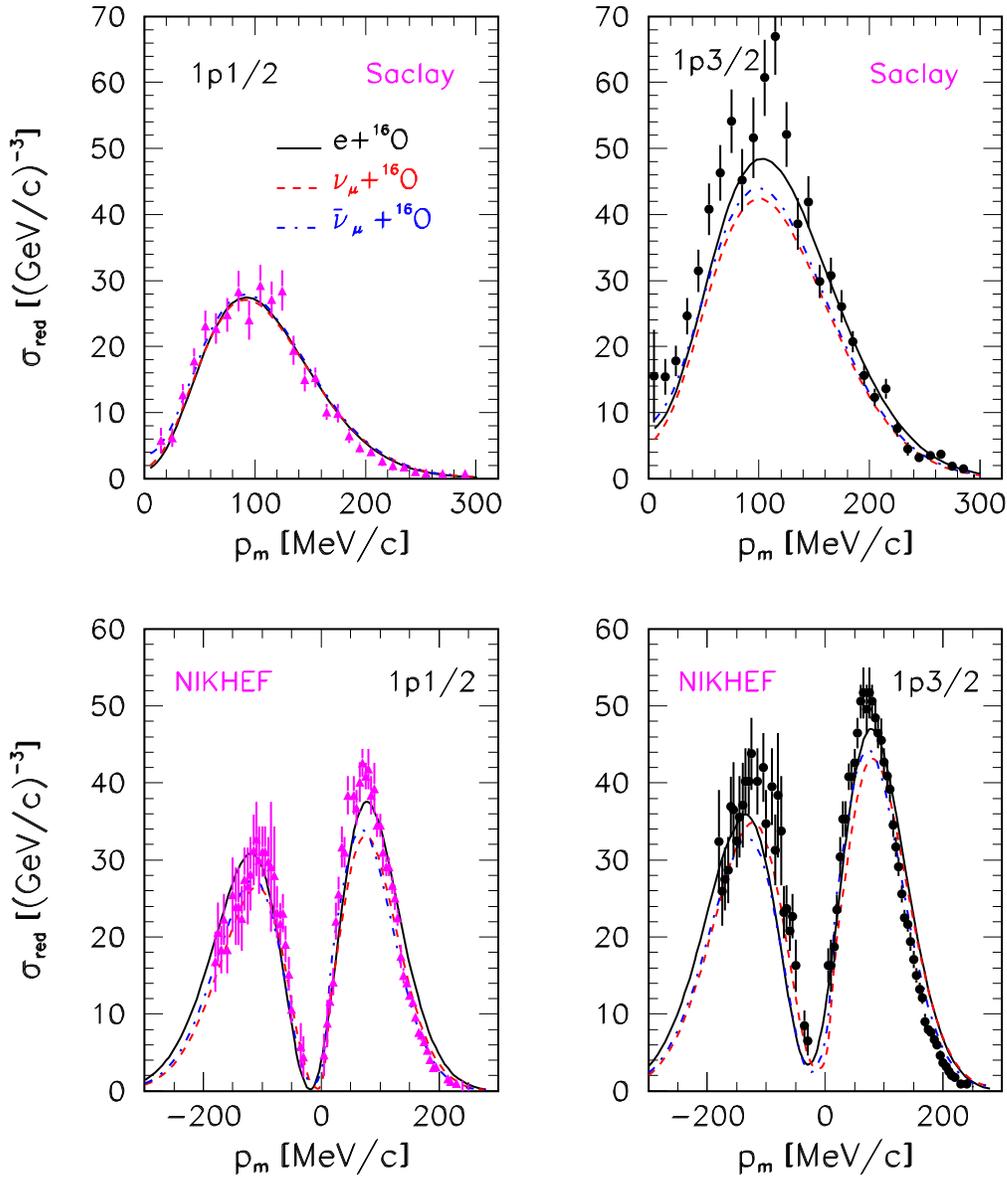}
  \end{center}
  \caption{(Color online) Comparison of the RDWIA electron, neutrino and
antineutrino reduced cross sections for the removal of nucleons from the 1$p$ shell
of $^{16}$O for Saclay \cite{REV37} and NIKHEF \cite{REV38}
kinematic as functions of $p_m$.
}
\end{figure*}
\begin{figure*}
  \begin{center}
    \includegraphics[height=40pc,width=40pc]{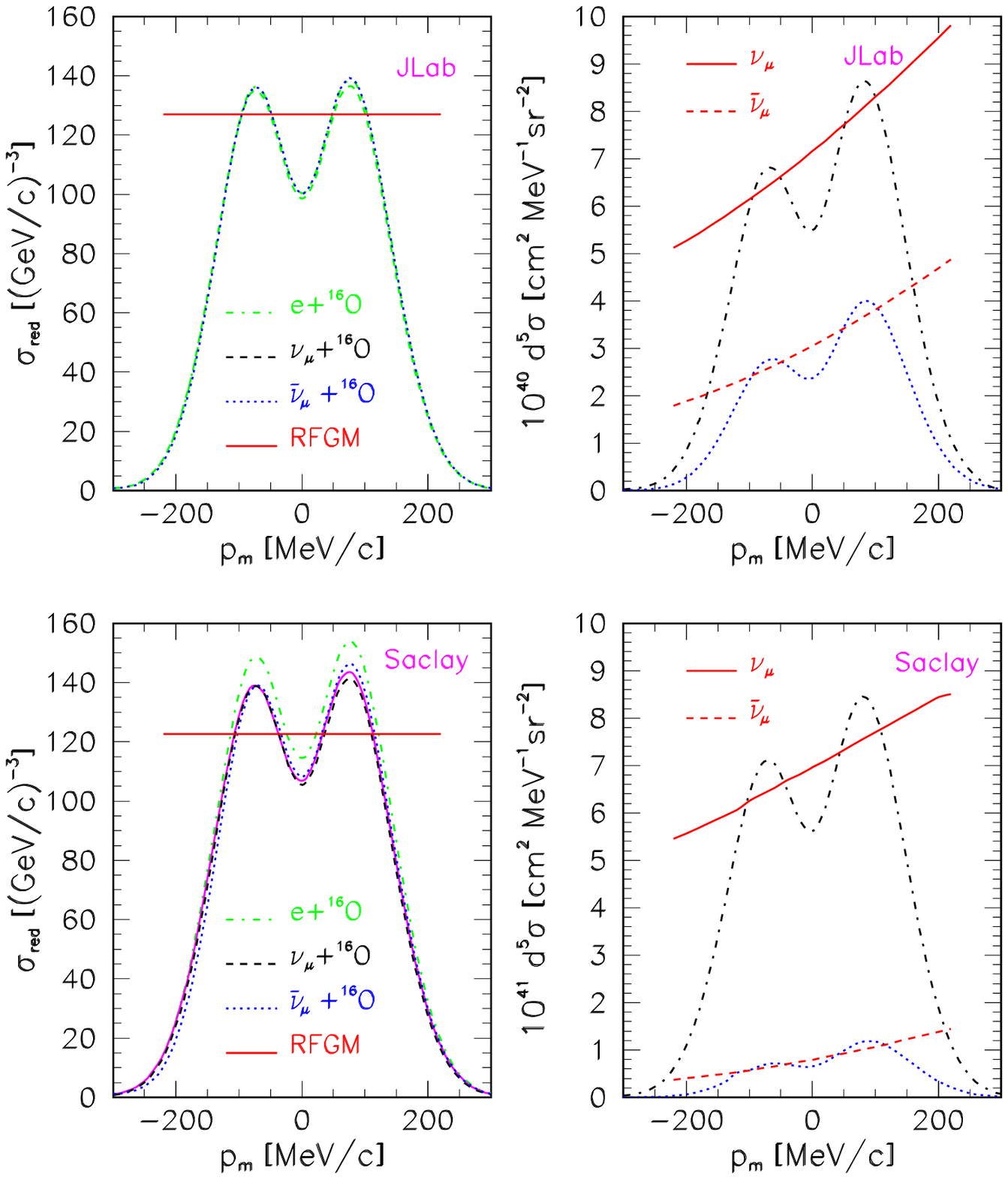}
  \end{center}
  \caption{(Color online) 
Comparison of the RDWIA and the RFGM calculations
for electron, neutrino and antineutrino reduced (left panels) and differential
(right panels) cross sections for the removal of nucleons from 1$p$ and
1$s$ shells of $^{16}$O as functions of missing momentum. 
The cross sections were calculated for the JLab \cite{REV19}
and Saclay \cite{REV36} kinematics. In the left panels, the RDWIA
calculations are shown for electron scattering (dashed-dotted line)
and neutrino (dashed line) and antineutrino (dotted line) scattering;
and the RFGM results are shown for the reduced cross sections (solid line).
In the right panels, the RFGM calculations are shown for the neutrino (solid line)
and antineutrino (dashed line) differential cross sections; and the RDWIA results
are shown for the neutrino (dashed-dotted line) and antineutrino (dotted line)
differential cross sections.
}
\end{figure*}

\section{Results}

The LEA code was successfully tested against $A(e,e^{\prime}p)$ data
\cite{REV18,REV19,REV20}. In Ref.~\cite{REV19} the uncertainty in the
normalization factors $S_{\alpha}$ was estimated to be about $\pm$15\%.
For illustration, Fig. 2 shows the measured JLab~\cite{REV19} and
Saclay~\cite{REV36} differential cross sections
for the removal of protons from the $1p$ shell of
$^{16}$O as functions of missing momentum $p_m$ as compared with LEA code
calculations. The reduced cross sections together with Saclay \cite{REV37}
and NIKHEF \cite{REV38} data are shown in Fig. 3. It should be noted that
negative values of $p_m$ correspond to $\phi=\pi$ and positive ones to
$\phi$=0. The cross sections were calculated
using the kinematic conditions with the normalization factors of data examined~\cite{REV19}.
Also shown in Figs. 2 and 3 are the results obtained in the PWIA
and RFGM (with the Fermi momentum $p_F$=225 MeV/c, binding
energy $\epsilon$=27 MeV and including the Pauli blocking factor). Apparently the
PWIA and RFGM overestimate the values of the cross sections, because the FSI
effects are neglected. Moreover, the RFGM predictions are completely off of the
exclusive data. This is because of the uniform momentum distribution of the
Fermi gas model.

The reduced cross sections for the removal of nucleons from $1p$ shell in
$^{16}$O$(e,e^{\prime}p)^{15}$N, $^{16}$O$(\nu,\mu^{-} p)^{15}$O, and
$^{16}$O$(\bar{\nu},\mu^{+} n)^{15}$N reactions are shown in Fig. 4 as
functions of $p_m$ together
with Saclay~\cite{REV37} and NIKHEF data. There is an overall good agreement
between calculated cross sections, but the value of electron cross sections
at the maximum is systematically higher (less than 10\%) than (anti)neutrino
ones with the exception of the $1p_{1/2}$ state for Saclay kinematics. The small
difference between neutrino and antineutrino reduced cross sections is
due to the difference in the FSI of proton and neutron with the residual
nucleus.

The differential and reduced electron and (anti)neutrino exclusive cross
sections for the removal of nucleons from $1p$  and $1s$ states were
calculated for JLab and Saclay~\cite{REV36} kinematics. The results are shown
in Fig. 5 together with the RFGM calculations. There is a good agreement
between all cross sections calculated in the RDWIA for JLab kinematics. The
difference between the electron and (anti)neutrino reduced cross sections
calculated for Saclay kinematics is less than 10\%. This can be attributed
to Coulomb distortion upon the electron wave function which is usually
described in the effective momentum approximation (EMA)~\cite{Schiff}.
In the EMA, the electron Coulomb wave function is replaced by a
plane wave with effective momentum whose value is larger than the value of
electron momentum at infinity, because of Coulomb attraction. This effect
weakens as the beam energy increases, and for this reason this effect is
more significant at Saclay kinematics ($E_{\rm beam}=500$ MeV) than at JLab
kinematics ($E_{\rm beam}=2442$ MeV).
Note that the RFGM results demonstrate absolutely different behavior.
\begin{figure*}
  \begin{center}
    \includegraphics[height=16cm,width=16cm]{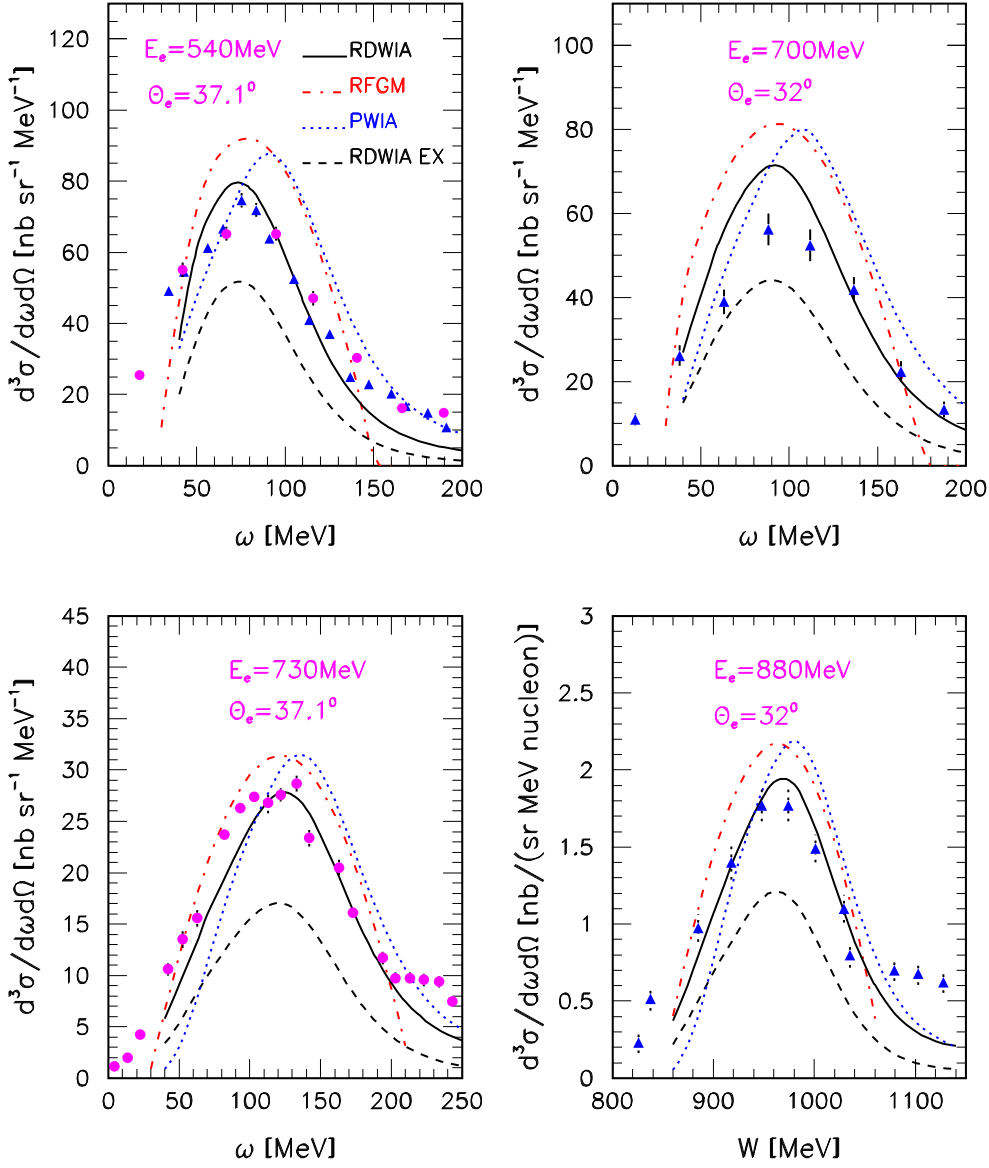}
  \end{center}
  \caption{(Color online) 
Inclusive cross section vs energy transfer
$\omega$ or invariant mass $W$ for electron scattering on
$^{16}$O. The data are from Ref.\cite{REV39} (SLAC, filled circles) and
Ref.\cite{REV40} (Frascati, filled triangles). SLAC data are for electron
beam energy $E_e$=540, 730 MeV and scattering angle
$\theta_e$=37.1$^{\circ}$. Frascati data are for $E_e$=540 MeV and
$\theta_e$=37.1$^\circ$, $E_e$=700, 880 MeV and $\theta_e$=32$^\circ$.
As shown in the key, cross sections were calculated with the RDWIA, PWIA, RFGM
and RDWIA with complex optical potential (EX).
}
\end{figure*}
\begin{figure*}
  \begin{center}
    \includegraphics[height=16cm,width=16cm]{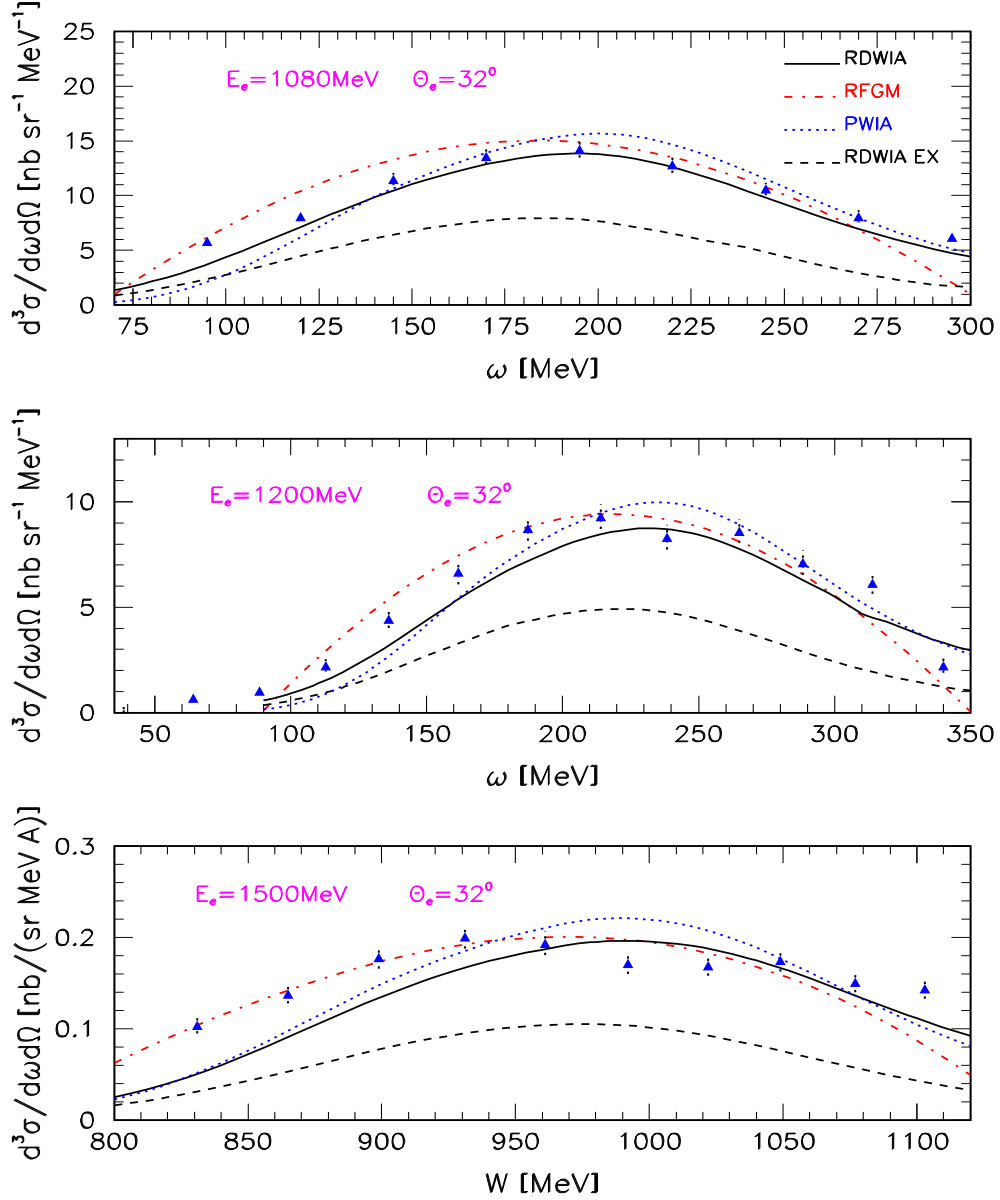}
  \end{center}
  \caption{(Color online) 
Same as Fig. 6, but the data are from Ref.\cite{REV40} for 
electron beam energy $E_e$=1080, 1200, and 1500 MeV and scattering angle
$\theta_e$=32$^{\circ}$.
}
\end{figure*}
\begin{figure}
  \begin{center}
    \includegraphics[height=16cm,width=16cm]{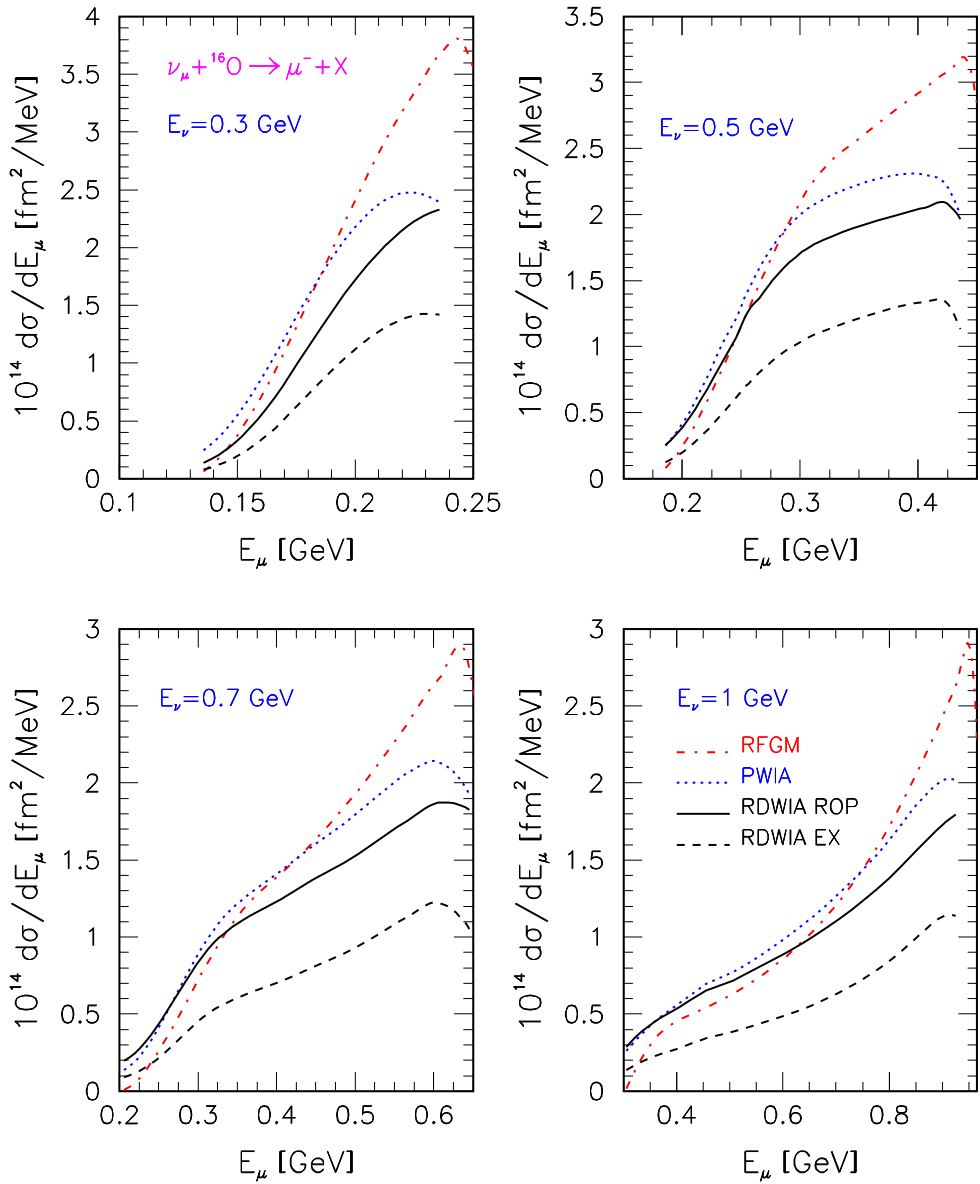}
  \end{center}
  \caption{(Color online) 
Inclusive cross section vs the muon energy for
neutrino scattering on $^{16}$O and for the four values of incoming neutrino
energy: $E_{\nu}$=0.3, 0.5, 0.7, and 1~GeV.
}
\end{figure}
\begin{figure}[t]
  \begin{center}
   \includegraphics[height=16cm, width=16cm ]{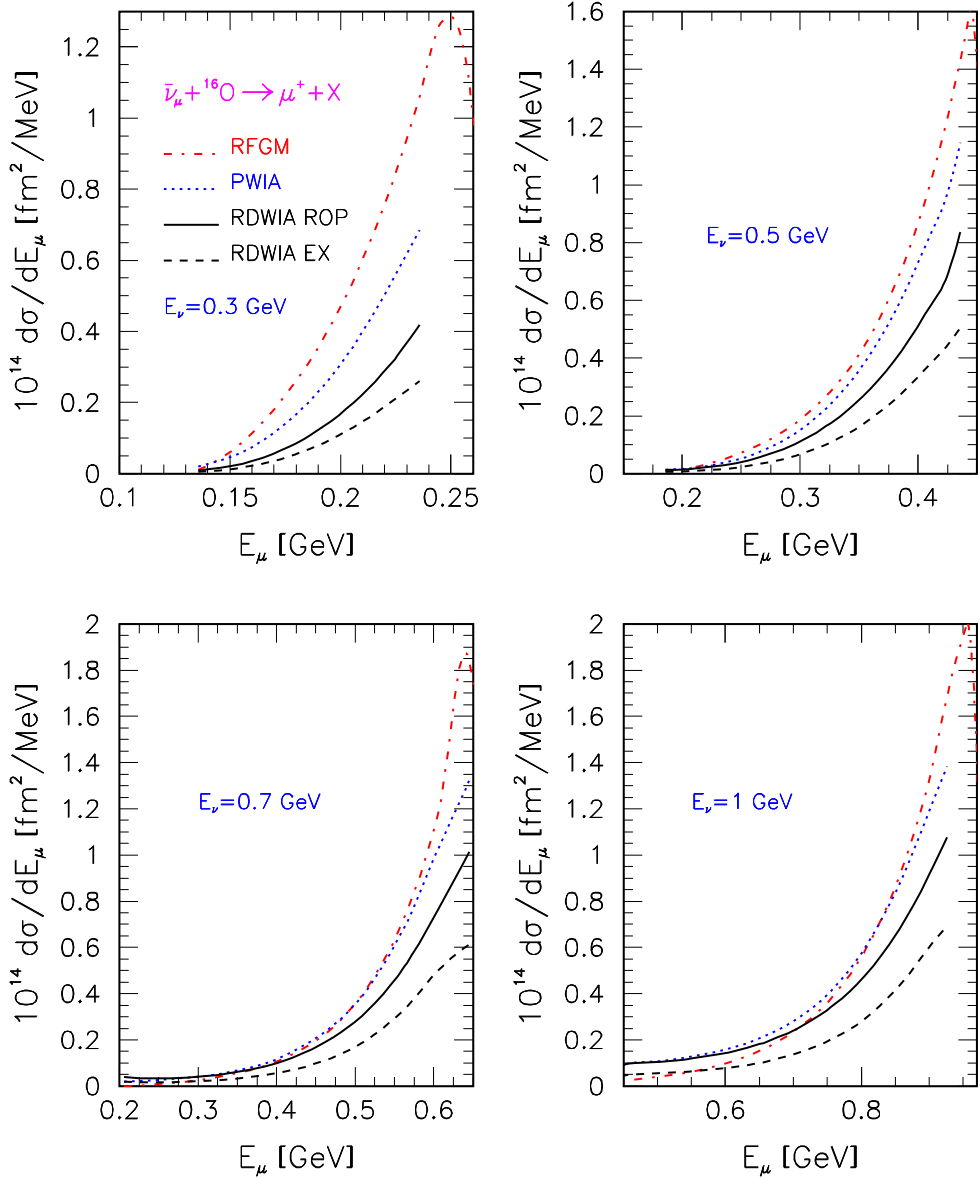}
  \end{center}
  \caption{(Color online) 
Same as Fig.~8, but for antineutrino scattering.
}
\end{figure}
\begin{figure}
  \begin{center}
    \includegraphics[height=16cm,width=16cm]{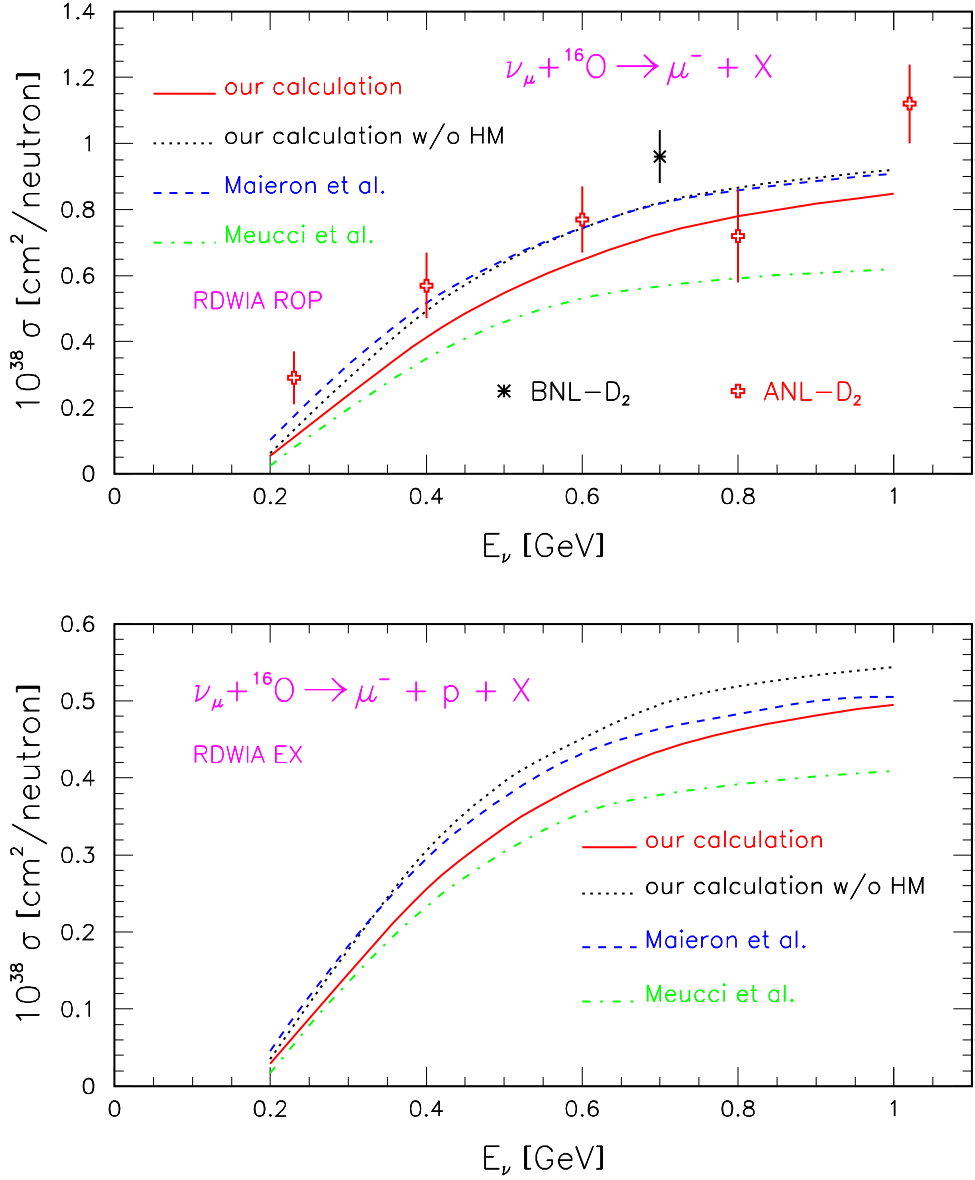}
  \end{center}
  \caption{(Color online) 
Total cross section for the CC QE scattering of muon
neutrino on $^{16}$O as a function of the incoming neutrino energy. The RDWIA
    results with the real part of optical potential (upper panel) and complex
    optical potential (lower panel) are shown together with calculations from
    Meucci {\it et al.} \cite{REV11} and Maieron {\it et al.} \cite{REV13}.
    The results obtained in this work were calculated
    with and without the contribution of the high-momentum component. For
    comparison, data for the  D$_2$ target are shown from Refs.\cite{REV41,REV42}.
}
\end{figure}
\begin{figure}
  \begin{center}
    \includegraphics[height=16cm,width=16cm]{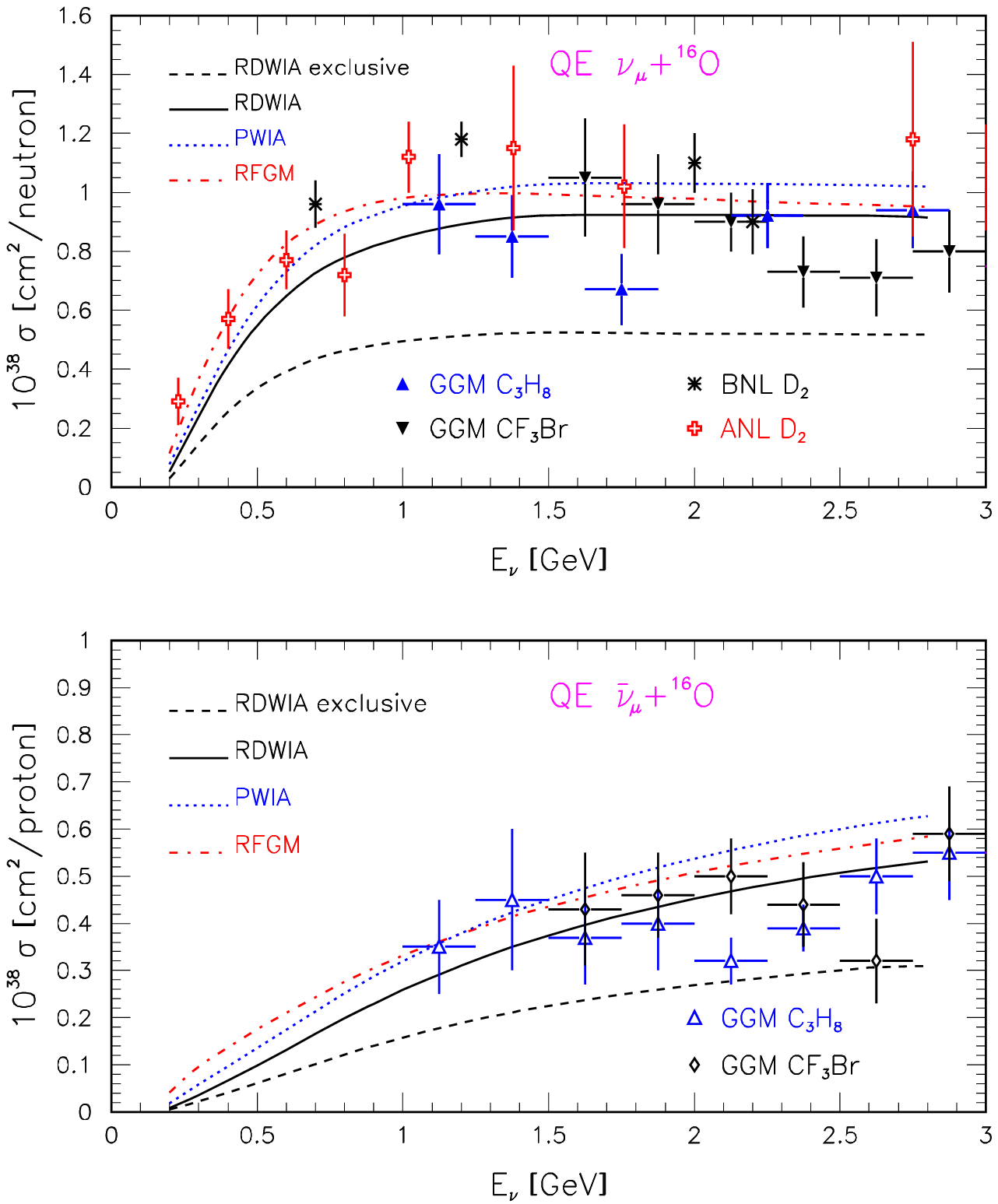}
  \end{center}
  \caption{(Color online) 
Total cross section for CC QE scattering of muon
neutrino (upper panel) and antineutrino (lower panel) on $^{16}$O as a
function of incoming (anti)neutrino energy.
Data points for different targets are from
Refs.\cite{REV41,REV42,REV43,REV44}.
}
\end{figure}

To test our approach, we calculated the inclusive
$^{16}$O$(e,e^{\prime})$ cross sections and compared them with SLAC
data~\cite{REV39} and Frascati data~\cite{REV40}. Figures 6 and 7 show
measured inclusive cross sections as functions of energy transfer, or the
invariant mass $W$ as compared with the RDWIA, PWIA, and RFGM calculations.
We note that relative to the PWIA results, the generic effect of the FSI with
the real part of the optical potential is to reduce the cross section value around
the peak and to shift the peak
toward the lower value energy transfer. The inclusion of the high-momentum
component increases the inclusive cross section in the high-energy transfer
region and improves the agreement with data. For the RDWIA results,
the difference between the calculated and measured cross sections at the
maximum are less than $\pm$10\%, with the exception of Frascati data for
$E_e=700$ MeV. For the RFGM results, these differences decrease with
$\vert\q\vert$ from about 22\% at $\vert\q\vert\approx330$ MeV/c down to
$\approx2$\% at $\vert\q\vert\approx640$ MeV/c. These results demonstrate
a strong  nuclear-model dependence of the inclusive cross sections at low
momentum transfer. This dependence weakens as $\vert\q\vert$ increases,
almost disappearing at $\vert\q\vert\geq 500$ MeV/c. The results for
$(e,e^{\prime}N)$ channel indicate that at least 50\% of the inclusive cross
section can be attributed to the single-step nucleon knockout.

The inclusive neutrino and antineutrino cross sections for energies
$E_{\nu}=300,\ 500,\ 700$, and 1000 MeV are presented in Figs. 8 and 9, which show
$d\sigma/dE_{\mu}$ as a function of muon energy. Here, the results obtained in
the RDWIA with the real optical potential (RDWIA ROP) are compared with the
inclusive cross sections calculated in the PWIA, RFGM, and RDWIA with complex
optical potential (RDWIA EX). The cross section values obtained in the RFGM
are higher than the ones obtained within the RDWIA ROP. For neutrino
(antineutrino) cross sections in the region close to the maximum, this
discrepancy is about 35\%(60\%) for $E_{\nu}=300$ MeV and 30\%(40\%) for
$E_{\nu}=1000$ MeV. The contribution of $(\nu,\mu N)$ channels to the
inclusive cross sections is about 60\%.

The total cross sections $\sigma(E_{\nu})$ together with data
~\cite{REV41,REV42} are presented in Fig. 10 as functions of the incident
neutrino energy. The upper panel shows the total cross sections for $^{16}$O
$(\nu,\mu^-)$ reaction calculated in the RDWIA with the real part of EDAD1
potential, and the lower panel shows the total cross sections for the
$^{16}$O$(\nu_{\mu},\mu^-p)$ channel. Also shown are the results obtained in
Refs.~\cite{REV11,REV13} with the NLSH bound nucleon wave functions, dipole
approximation of the nucleon form factors, EDAD1 optical potential and
neglecting the $NN$ correlation  contributions. The cross sections are scaled
with the number of neutrons in the target.

Our values of  $(\nu,\mu^-)$[$(\nu_{\mu},\mu^-p)$] cross sections are
systematically larger than those from Ref.~\cite{REV11}. The discrepancy
increases with energy from about 17\%(7\%) for $E_{\nu}=300$ MeV up to
28\%(20\%) for $E_{\nu}=1000$ MeV.
On the other hand, our cross sections are lower than those from
Ref.~\cite{REV13}, and the discrepancy decreases with energy from 37\%(15\%)
for $E_{\nu}=300$ MeV upto 15\% (7\%) down to $E_{\nu}=1000$ MeV. 
To study the $NN$ correlation  effect, we calculated the total cross sections
without the high-momentum contribution, i.e., with $S_{\alpha}=1$ for all bound
nucleon states, similar to Refs~\cite{REV11,REV13}.
The results are shown in Fig. 10. Apparently, the $NN$ correlation effect reduces
the total cross section. The difference between the results obtained with and
without the high-momentum component contribution decreases with neutrino
energy from about 20\% for $E_{\nu}=200$ MeV down to $\approx8$\% for
$E_{\nu}=1000$ MeV. Moreover, in this case the agreement with the result of
Ref.\cite{REV13} is good, and the discrepancy is less then
$\pm6$\% for $E_{\nu}>300$ MeV.

The neutrino and antineutrino total cross sections calculated up to
neutrino energy 2.5 GeV are shown in Fig. 11 together with data of 
Refs.\cite{REV41,REV42,REV43,REV44}. Also shown are the results obtained in the RFGM
and PWIA as well as the contribution of the exclusive channels to the total cross
sections. The cross sections are scaled with the neutron/proton number 
in the target. The ratio between the neutrino cross sections calculated in the
RFGM and RDWIA ROP decreases with neutrino energy from about 1.5 for
$E_{\nu}=300$ MeV to $\approx1.18$ for $E_{\nu}=1$ GeV and down to
$\approx1.05$ for $E_{\nu}=2.4$ GeV. For the antineutrino cross sections, this
ratio is about 2.7 for $E_{\nu}=300$ MeV, 1.3 for $E_{\nu}=1$ GeV,
and 1.1 for $E_{\nu}=2.4$ GeV.

It follows from the comparison of the PWIA and RDWIA results that the FSI effects
reduce the total cross section. For the neutrino interactions,
 this reduction is about 16\% for $E_{\nu}=300$ MeV and decreases
slowly to 10\% for $E_{\nu}=2.4$ GeV. The reduction of the
antineutrino cross section is about 38\% for $E_{\nu}=300$ MeV 
and $\approx15$\% for $E_{\nu}=2.4$ GeV.
We, therefore, observe the weakening of FSI effect in total cross sections
with the increase of energy transfer, in accordance with the calculation of
Ref.\cite{REV45}. The contribution of the exclusive channels is
about 60\%. The results presented in Fig. 11 show significant nuclear-model
dependence for energy less than 1 GeV.

\section{Conclusions}

In this paper, we study electron and CC quasi-elastic (anti)neutrino
scattering on the oxygen target in different approximations (PWIA, RDWIA,
RFGM) placing particular emphasis on the nuclear-model dependence of the
results.
In RDWIA, the LEA program, adapted to neutrino interactions, was used to
calculate the differential and reduced exclusive cross sections.
This approach was earlier applied to electron-nucleus
scattering and successfully tested against data.
We found that the reduced cross sections for (anti)neutrino scattering are
similar to those of electron scattering, and the latter are in a good
agreement with electron data.
In calculating the inclusive and total cross sections, the imaginary
part of relativistic optical potential was neglected and the effect of
$NN$ correlations in the target ground state was taken into account. This
approach was tested against electron-oxygen inclusive scattering data; there was
overall agreement with the data, with the differences between
calculated and measured cross sections in the peak region less than 10\%.
For neutrino interactions the FSI effect reduces the total cross
section by about 30\% for $E_{\nu}$=200 MeV compared to PWIA and
decreases with neutrino energy down to 10\% at 1 GeV. The effect of
$NN$ correlations reduces the total cross section by about 15\% at
$E_{\nu}$=200 MeV and also decreases with neutrino energy down to about
8\% at 1 GeV.

We tested the RFGM against electron-oxygen scattering data and found
that this model does not reproduce the exclusive cross section
data. The RFGM also leads to an overestimated value of the inclusive
$^{16}$O$(e,e^{\prime})$ cross section at low momentum transfer. The
discrepancy is about 20\% and decreases as momentum transfer increases.
The values of the (anti)neutrino cross sections calculated in this model
are also higher than the corresponding values in the RDWIA approach.

We conclude that the data favor the RDWIA results. 
This indicates that the use of RDWIA in Monte Carlo simulations
of neutrino detector response would allow one to reduce the systematic
uncertainty in neutrino oscillation parameters.

\section*{Acknowledgments}

The authors greatly acknowledge communications with J. J. Kelly whose LEA
code for nucleon knockout by electron scattering was adapted in this work
for neutrino interactions.
This work was partially supported by Russian Foundation for Basic
Research, Project Nos. 06-02-16353, 06-02-16659, and 05-02-17196.

\appendix

\section{Hadronic tensor and cross section of exclusive electron and
neutrino scattering}
\label{A}

A general structure of the hadronic tensor can be derived from
the requirements of Lorentz invariance, parity, and time
reversal symmetries. For unpolarized nucleon and a nucleus in the final
state, this tensor must be constructed from three linearly
independed four-vectors $q$, $p_x$, and $p_A$, the scalars that can be
constructed from them, the second-rank metric tensor $g_{\mu \nu}$, and
completely antisymmetric tensor $\epsilon_{\mu \nu \alpha \beta}$. Generally,
because of the final state interaction effects, the scattered flux at infinity
involves complicated asymptotic configurations, and the time reversal symmetry
does not constraint the form of the nuclear tensor for the
exclusive reactions ~\cite{REV23,REV24,DON}.

\subsection{Electron scattering}

For electron scattering, the
electromagnetic current conservation requires
$q_{\mu}W^{\mu \nu}=W^{\mu\nu}q_{\nu}=0$. Taking into account the
parity conservation, the nuclear tensor can then be written as a sum of
symmetric, $W_S^{\mu \nu (el)}$, and
antisymmetric, $W_A^{\mu \nu (el)}$, parts \cite{REV25}
\begin{widetext}
\begin{subequations}
\begin{align}
W^{\mu \nu (el)} &=W_S^{\mu \nu (el)} + W_A^{\mu \nu (el)},     
\\
W_S^{\mu \nu (el)} &= W_1^{(el)}\tilde{g}^{\mu \nu}
+ W_2^{(el)}\tilde{p}_x^{\mu}\tilde{p}_x^{\nu}
+ W_3^{(el)}\tilde{p}_A^{\mu}\tilde{p}_A^{\nu}
 + W_4^{(el)}(\tilde{p}_x^{\mu}\tilde{p}_A^{\nu}+
\tilde{p}_x^{\nu}\tilde{p}_A^{\mu}),
\\
W_A^{\mu \nu (el)} &=W_5^{(el)}(\tilde{p}^{\mu}_x\tilde{p}^{\nu}_A -
\tilde{p}^{\nu}_x\tilde{p}^{\mu}_A),
\end{align}
\end{subequations}
\end{widetext}
where
\begin{subequations}
\begin{align}
\tilde{g}^{\mu \nu}=g^{\mu \nu} + \frac{q^{\mu}q^{\nu}}{Q^2},    
\\
\tilde{p}_x^{\mu}=p_x^{\mu} +\frac{p_x\cdot q}{Q^2}q^{\mu},
\\
\tilde{p}_A^{\mu}=p_A^{\mu} +\frac{p_A\cdot q}{Q^2}q^{\mu}.
\end{align}
\end{subequations}
In target rest frame, the coordinate system is chosen such that the {\it z} axis
is parallel to the momentum transfer ${\bf q}={\bf k}_i-{\bf k}_f$ and the
{\it y} axis is parallel to ${\bf k}_i \times {\bf k}_f$, and the components of
the four-vectors are
$k_f=(\varepsilon_f,\vert {\bf k}_f \vert\sin\theta\cos\varphi,
\vert {\bf k}_f \vert\sin\theta\sin\varphi,\vert {\bf k}_f \vert\cos\theta)$,
 $q=(\omega,0,0,\vert {\bf q} \vert)$, $p_A=(m_A,0,0,0)$,
$p_x=(\varepsilon_x,\vert {\bf p}_x \vert\sin \theta_{x}\cos\phi,
\vert {\bf p}_x \vert\sin \theta_{x}\sin\phi,
\vert {\bf p}_x \vert\cos \theta_{x})$,
where $\theta$, $\varphi$ are lepton scattering angles and $\theta_{x}$,
$\phi$ are the outgoing nucleon angles.

The lepton tensor for unpolarized electron scattering is symmetric, and
therefore the result of contraction of the electron and nuclear response
tensors reduces to the form
\begin{widetext}
\begin{eqnarray}
L^{(el)}_{\mu \nu}W_S^{\mu \nu (el)} &=& 4\varepsilon_i \varepsilon_f
\cos^2\frac{\theta}{2}(V_LR_L^{(el)}   
 + V_TR_T^{(el)} + V_{LT}R_{LT}^{(el)}\cos\phi   
 + V_{TT}R_{TT}^{(el)}\cos 2\phi ),
\end{eqnarray}
\end{widetext}
where
\begin{subequations}
\begin{align}
V_L &={Q^4}/{\q^4 },
\\
V_T &=\frac{Q^2}{2\q^2 } + \tan^2\frac{\theta}{2},
\\
V_{LT} &=\frac{Q^2}{ \q^2}\left(\frac{Q^2}{ \q^2 }                      
 + \tan^2\frac{\theta}{2}\right)^{1/2},
\\
V_{TT} &=\frac{Q^2}{2\q^2 },
\end{align}
\end{subequations}
are the electron coupling coefficients, and
\begin{subequations}
\begin{align}
R^{(el)}_L &=W^{00 (el)},
\\
R^{(el)}_T &=W^{xx (el)}+W^{yy (el)},                                  
\\
R^{(el)}_{LT}\cos\phi &=-\left(W^{0x (el)}+W^{x0 (el)}\right),
\\
R^{(el)}_{TT}\cos2\phi &=W^{xx (el)} - W^{yy (el)},
\end{align}
\end{subequations}
are four independ response functions, which describe the electromagnetic
properties of the hadronic system.

\subsection{Neutrino scattering}

In weak interactions, the weak current and parity are not conserved.
Therefore, a general nuclear tensor can be written as
\begin{widetext}
\begin{subequations}
\begin{align}
W^{\mu \nu (cc)} = & \phantom{+}W^{\mu \nu}_S + W^{\mu \nu}_A,
\\
W^{\mu \nu}_S = & \phantom{+}W_1 g^{\mu \nu} + W_2 q^{\mu}q^{\nu}
                + W_3 p^{\mu}_xp^{\nu}_x + W_4 p^{\mu}_A p^{\nu}_A
  + W_5 (p^{\mu}_xq^{\nu} +p_x^{\nu}q^{\mu})
\notag \\
&
{} + W_6 (p^{\mu}_Aq^{\nu} + p^{\nu}_Aq^{\mu})
 + W_7 (p_x^{\mu}p_A^{\nu}+p_x^{\nu}p_A^{\mu}),
\\                                                                      
W^{\mu \nu}_A = & \phantom{+}W_8(p^{\mu}_xq^{\nu} -p_x^{\nu}q^{\mu})
+ W_9(p^{\mu}_Aq^{\nu} - p^{\nu}_Aq^{\mu})            
+ W_{10}(p_x^{\mu}p_A^{\nu}-p_x^{\nu}p_A^{\mu})
\notag \\
&
{} + W_{11}\epsilon^{\mu\nu\tau\rho}q_{\tau}p_{x\rho}
   + W_{12}\epsilon^{\mu\nu\tau\rho}q_{\tau}p_{A\rho}
   + W_{13}\epsilon^{\mu\nu\tau\rho}p_{x\tau}p_{A\rho} .
\end{align}
\end{subequations}
\end{widetext}
Note that because of Hermicity of $W^{\mu\nu(cc)}$, each term of
$W^{\mu\nu}_S$ must be real, while each term of $W^{\mu\nu}_A$ must be
imaginary, and $L^{cc}_{\mu \nu}W^{\mu \nu (cc)}$ is real.
The result of contraction of the lepton and nuclear tensors can be written as
\begin{widetext}
\begin{align}
L_{\mu \nu}^{(cc)}W^{\mu \nu (cc)} = &\phantom{+}L^S_{\mu \nu}W^{\mu \nu}_S +
L^A_{\mu \nu}W^{\mu \nu }_A =
2\varepsilon_i\varepsilon_f
\{v_0R_0 + v_TR_T
+ v_{TT}R_{TT}\cos 2\phi                                   
+ v_{zz}R_{zz}
\notag \\
&
+ (v_{xz}R_{xz} - v_{0x}R_{0x})\cos\phi
-v_{0z}R_{0z}
+ h[ v_{yz}(R^{\prime}_{yz}\sin\phi + R_{yz}\cos\phi)
\notag \\
&
- v_{0y}(R^{\prime}_{0y}\sin\phi +
R_{0y}\cos\phi) - v_{xy}R_{xy}]\},
\end{align}
\end{widetext}
where
\begin{subequations}
\begin{align}
v_0 & = 1+\beta \cos \theta, \\
v_T & = 1-\beta\cos \theta +
\frac{\varepsilon_i\beta\vert \k_f\vert
\sin^2 \theta}{ \ q^2}, \\
v_{TT} & = \frac{\varepsilon_i\beta\vert \k_f\vert
\sin^2 \theta}{ \q^2}, \\
v_{0z}&  = \frac{\omega}{\vert\q\vert}(1 + \beta\cos \theta)+
\frac{m^2_l}{\vert \q\vert \varepsilon _f}, \\
v_{zz} & = 1 + \beta\cos \theta -2\frac
{\varepsilon _i \vert \k_f \vert \beta}{ \q ^2}\sin^2 \theta, \\
v_{0x} & = (\varepsilon_i+\varepsilon_f)\frac{\beta \sin \theta}       
{\vert\q\vert}, \\
v_{xz} & = \frac{\beta}{\q^2}\sin\theta
\left[(\varepsilon_i+\varepsilon_f)\omega+m_l^2)\right], \\
v_{xy} & = \frac{\varepsilon_i+\varepsilon_f}{\vert \q\vert }(1-\beta
\cos\theta) -\frac{m^2_l}{\vert \q\vert \varepsilon_f}, \\
v_{yz} & = \beta \frac{\omega}{\vert\q\vert} \sin \theta,\\
v_{0y} & = \beta \sin\theta, ~~~
\beta =\vert\k_f\vert/\varepsilon_f ,
\end{align}
\end{subequations}
are neutrino coupling coefficients, and
\begin{subequations}
\begin{align}
R_0 & = W_S^{00},\\
R_T & = W_S^{xx}+W_S^{yy},\\
R_{TT}\cos2\phi & = W_S^{xx}-W_S^{yy},\\
R_{0z}&  = W_S^{0z}+W_S^{z0},\\
R_{zz} & = W_S^{zz}, \\                                              
R_{0x}\cos\phi & = W_S^{0x}+W_S^{x0}, \\
R_{xz}\cos\phi & = W_S^{xz}+W_S^{zx}, \\
R_{xy} & =  i\left(W_A^{xy}-W_A^{yx}\right), \\
R^{\prime}_{yz}\sin\phi + R_{yz}\cos\phi & = i\left(W_A^{yz}-W_A^{zy}\right),\\
R^{\prime}_{0y}\sin \phi+R_{0y}\cos \phi & = i\left(W_A^{0y}-W_A^{y0}\right),
\end{align}
\end{subequations}
are ten independ response functions which describe the weak properties of
the hadronic system.

In the absence of FSI effect (plane-wave limit) the nucleon flux conserves in
exclusive reaction. For this reason, the time reversal symmetry of operators
and states provides an additional constraint on the Lorenz form
of the antisymmetric part of nuclear tensor (A6c), in particular, the
structures like $a^{\mu}b^{\nu}-a^{\nu}b^{\mu}$ (with $a$ and $b$
four-momenta) vanish. Then we have
\begin{widetext}
\begin{eqnarray}
W^{\mu \nu}_A &=& W_{11}\varepsilon^{\mu\nu\tau\rho}q_{\tau}p_{x\rho}
 + W_{12}\epsilon^{\mu\nu\tau\rho}q_{\tau}p_{A\rho}      
+ W_{13}\epsilon^{\mu\nu\tau\rho}p_{x\tau}p_{A\rho}                   
\end{eqnarray}
and
\begin{eqnarray}
L^{(cc)}_{\mu\nu}W^{\mu \nu (cc)}&= 2\varepsilon_i\varepsilon_f
\{v_0R_0 + v_TR_T
 + v_{TT}R_{TT}\cos 2\phi                                
 + v_{zz}R_{zz}
 + (v_{xz}R_{xz} - v_{0x}R_{0x})\cos\phi
\nn \\
&
{} - v_{0z}R_{0z}
 + h( v_{yz}R_{yz}\cos\phi - v_{0y}R_{0y}\cos\phi       
 - v_{xy}R_{xy})\},
\end{eqnarray}
\end{widetext}
where
\begin{subequations}
\begin{align}
R_{yz}\cos\phi & = i\left(W_A^{yz}-W_A^{zy}\right), \\
R_{0y}\cos \phi & = i\left(W_A^{0y}-W_A^{y0}\right).                   
\end{align}
\end{subequations}
Note that the response functions $R^{\prime}_{yz}$ and $R^{\prime}_{0y}$ are
related to $W_{8-10}$ terms which vanish in the plane-wave limit.
It follows from the expressions (A7) and (A11) that the cross sections
asymmetry, which is measured at azimuthal angles $\phi=\pi/2$ and
$\phi=-\pi/2$, vanishes in the absence of the FSI.



\begin{thebibliography}{99}

\bibitem{REV1} S.~Hatakeyama {\it et al.} (Super-Kamiokande Collaboration),
Phys. Rev. Lett. {\bf 81}, 2016, 1998; M.~Ambrosio {\it et al.} (MACRO
Collaboration), Phys. Lett. {\bf B434}, 451, 1998; W.~W.~Allison {\it et al.}
(Soudan-2 Collaboration), Phys. Lett. {\bf B449}, 137, 1999.

\bibitem{REV2} Y.~Fukuda {\it et al.} (Super-Kamiokande Collaboration),
Phys. Rev. Lett. {\bf 81}, 1562, 1998; Erratum-ibid. {\bf 81}, 4279, 1998;
B.~T.~Cleveland {\it et al.}, Astrophys. J. {\bf 496}, 505, 1998, W.~Hamplet
{\it et al.} (GALLEX Collaboration), Phys. Lett. {\bf B447}, 127, 1999;
J.~N.~Abdurashitov {\it et al.} (SAGE Collaboration), Phys. Rev. {\bf C60},
055801, 1999.

\bibitem{REV3} B.~Aharmim {\it et al.} (SNO Collaboration), Phys.
Rev. {\bf C72}, 055502, 2005;
Y.~Ashie {\it et al.} (Super-Kamiokande Collaboration), Phys. Rev.
 {\bf D71}, 112005, 2005.

\bibitem{REV4} T.~Araki {\it et al.}, Phys. Rev. Lett. {\bf 94}, 081801, 2005.

\bibitem{REV5} M.~H.~Ahn {\it et al.}, Phys. Rev. {\bf D74}, 072003, 2006.

\bibitem{REV6} D.~G.~Michael {\it et al.}, Phys. Rev. Lett. {\bf 97}, 191801,
2006.

\bibitem{REV7}G.~P.~Zeller, arXiv:hep-ex/0312061.

\bibitem{REV8} A.~V.~Butkevich and S.~P.~Mikheyev, Phys. Rev. {\bf C72},
  025501, 2005.

\bibitem{REV9} {\it Modern Topics in Electron Scattering}, edited by B.~Frois
and I.~Sick (World Scientific, Singapore, 1991).

\bibitem{REV10}O.~Benhar and D.~Meloni, Nucl. Phys. {\bf A789}, 379 (2007). 

\bibitem{REV11} A.~Meucci, C.~Giusti, and F.~D.~Pacati, Nucl. Phys.
{\bf A739}, 277, 2004.

\bibitem{REV12} A.~Meucci, C.~Giusti, and F.~D.~Pacati, Nucl. Phys.
{\bf A744}, 307, 2004.

\bibitem{REV13} C.~Maieron, M.~C.~Martinez, J.~A.~Caballero, and J.~M.~Udias,
Phys. Rev. {\bf C68}, 048501, 2003.

\bibitem{REV14} M.~C.~Martinez, P.~Lava, N.~Jachowicz, J.~Ryckebusch,
  K.~Vantournhout, and J.~M.~Udias, Phys. Rev. {\bf C73}, 024607, 2006.

\bibitem{REV15} J.~Nieves, J.~E.~Amaro, and M.~Valverde, Phys. Rev. {\bf C70},
055503, 2004.

\bibitem{Kolb} E.~Kolbe, K.~Langanke, S.~Krewald, F.~K.~Thielemann, Nucl. Phys.
{\bf A540}, 599, 1992.

\bibitem{Vopl} C.~Vople, N.~Auerbach, G.~Colo, T.~Suzuki, N.~Van~Giani,
Phys. Rev. {\bf C62}, 015501, 2000.

\bibitem{Ryck} N.~Jachowicz, K.~Heyde, J.~Ryckebusch, S.~Rombouts,
Phys. Rev. {\bf C65}, 025501, 2002.

\bibitem{Singh} S.~K.~Singh, Nimai~C.~Mukhopadhyay, E.~Oset, Phys. Rev.
{\bf C57}, 2687, 1998.

\bibitem{REV16} J.~J~Kelly, http://www.physics.umd.edu/enp/jjkelly/LEA

\bibitem{REV17} J.~J~Kelly, Phys. Rev. {\bf C72}, 014602, 2005

\bibitem{REV18} J.~Gao {\it et al.}, Phys. Rev. Lett. {\bf 84}, 3265, 2000

\bibitem{REV19} K.~G.~Fissum {\it et al.}, Phys. Rev. {\bf C70}, 034606, 2004

\bibitem{REV20} J.~J.~Kelly, Phys. Rev. {\bf C71}, 064610, 2005

\bibitem{Frull} S.~Frullani and J.~Mougey, Adv. Nucl. Phys. {\bf 14}, 1 (1984).


\bibitem{REV21} C.~Ciofi degli Atti and S.~Simula, Phys. Rev. {\bf C53 },
1689, 1996.

\bibitem{REV22} S.~A.~Kulagin and R.~Petti, Nucl. Phys. {\bf A765}, 126, 2006.

\bibitem{REV23} A.~Picklesimer, J.~W.~Van Orden, S.~J.~Wallace,
Phys. Rev. {\bf C32}, 1312, 1985.

\bibitem{REV24} A.~Picklesimer, J.~W.~Van Orden, Phys. Rev.
{\bf C35}, 266, 1987.

\bibitem{DON} T.~W.~Donnelly and A.~S.~Raskin, Ann. Phys.\  
{\bf 169}, 247, 1986.

\bibitem{REV25} J.~J.~Kelly, Adv. Nucl. Phys. {\bf 23}, 75, 1996.

\bibitem{REV26} T.~de~Forest, Nucl. Phys. {\bf A392}, 232, 1983.

\bibitem{REV27} P.~Mergell, U.-G.~Meissner, and D.~Drechesel, Nucl. Phys.
{\bf A596}, 367, 1996.

\bibitem{REV28} J.~J.~Kelly, Phys. Rev. {\bf C59}, 3256, 1996

\bibitem{REV29} C.~Horowitz and B.~Serot, Nucl. Phys. {\bf A368}, 503, 1981.

\bibitem{REV30} M.~M.~Sharma, M.~A.~Nagarajan, and P.~Ring, Phys. Lett.
{\bf B312}, 377, 1993.

\bibitem{REV31} M.~Leuschner {\it et al.}, Phys. Rev. {\bf C49}, 955, 1994

\bibitem{REV32} J.~M.~Udias, P.~Sarriguren, E.~Moya de Guerra, E.~Garrido, and
J.~A.~Caballero, Phys. Rev. {\bf C51}, 3246, 1995.

\bibitem{REV33} M.~Hedayati-Poor, J.~I.~Johansson, and H.~S.~Sherif,
Phys. Rev. {\bf C51}, 2044, 1995.

\bibitem{REV34} E~.D.~Cooper, S.~Hama, B.~C.~Clark, and R.~L.~Mercer,
Phys. Rev. {\bf C47}, 297, 1993.

\bibitem{REV35} A.~Meucci, F.~Capuzzi, C.~Giusti, and F.~D.~Pacati,
Phys. Rev. {\bf C67}, 054601, 2003.

\bibitem{REV36} L.~Chinitz {\it et al.}, Phys. Rev. Lett.{\bf 67}, 568, 1991.

\bibitem{REV37} M.~Bernhein {\it et al.}, Nucl. Phys.{\bf A375}, 381, 1982.

\bibitem{REV38} M.~Leuschner {\it et al.}, Phys. Rev. {\bf C49}, 955, 1994.

\bibitem{Schiff} L.~L.~Schiff Phys.Rev. {\bf103}, 443, 1956.

\bibitem{REV39} J.~S.~O'Connell {\it et al.}, Phys. Rev. {\bf C35}, 1063, 1987.

\bibitem{REV40} M.~Anghinolfi {\it et al.}, Nucl. Phys.{\bf A602}, 402, 1996.

\bibitem{REV41} W.~A.~Mann {\it et al.}, Phys. Rev. Lett.{\bf 31}, 844, 1973.

\bibitem{REV42} N.~J.~Baker {\it et al.}, Phys. Rev. {\bf D23}, 2499, 1981.

\bibitem{REV43} M.~Pohl {\it et al.}, Lett. Nuovo Cim. {\bf 26}, 332, 1979.

\bibitem{REV44} J.~Brunner {\it et al.}, Z. Phys. {\bf C45}, 551, 1990.

\bibitem{REV45} Y.~Horikawa, F.~Lenz and Nimai~C.~Mukhopadhyay, Phys. Rev.
{\bf C22}, 1680, 1980.

\end{thebibliography}
\end{document}